\documentclass[%
 reprint,
superscriptaddress,
 amssymb,
 aps,
 pra,
 sort&compress,numbers,
]{revtex4-2}

\usepackage{graphicx}
\usepackage{dcolumn}
\usepackage{braket}
\usepackage{amsmath}
\usepackage{verbatim}
\usepackage{cprotect}
\usepackage{soul}
\usepackage{bm}
\usepackage[colorlinks=True]{hyperref}
\newif\ifistoreview

\istoreviewtrue
\newcommand{\setreviewsoff}{\istoreviewfalse}
\setreviewsoff
\usepackage{soul}
\usepackage{xcolor}

\newcommand{\mc}[1]{\mathcal{#1}}

\begin{document}
\preprint{APS/123-QED}

\title{Direct experimental observation of sub-poissonian photon statistics by means of multi-photon scattering on a two-level system}
\author{A.Yu. Dmitriev}
\email{aleksei.j.dmitriev@phystech.edu}
\affiliation{Laboratory of Artificial Quantum Systems, \\ Moscow Institute of Physics and Technology, 141700 Dolgoprudny, Russia}

\author{A.V. Vasenin}%
\affiliation{Laboratory of Artificial Quantum Systems, \\ Moscow Institute of Physics and Technology, 141700 Dolgoprudny, Russia}
\affiliation{Skolkovo Institute of Science and Technology, Nobel St. 3, 143026 Moscow, Russia}

\author{S.A. Gunin}
\affiliation{Laboratory of Artificial Quantum Systems, \\ Moscow Institute of Physics and Technology, 141700 Dolgoprudny, Russia}
\affiliation{Skolkovo Institute of Science and Technology, Nobel St. 3, 143026 Moscow, Russia}
\author{S.V.~Remizov}
\affiliation{Dukhov Research Institute of Automatics (VNIIA), 127055 Moscow, Russia}
  \affiliation{V. A. Kotel'nikov Institute of Radio Engineering and Electronics, Russian Academy of Sciences, Moscow 125009, Russia}
  \affiliation{HSE University, Moscow 109028, Russia}
\author{A.A. Elistratov}
\affiliation{Dukhov Research Institute of Automatics (VNIIA), 127055 Moscow, Russia}
\author{W.V. Pogosov}
\affiliation{Dukhov Research Institute of Automatics (VNIIA), 127055 Moscow, Russia}
\affiliation{Advanced Mesoscience and Nanotechnology Centre, Moscow Institute of Physics and Technology, Dolgoprudny, 141700, Russia}
\affiliation{Institute for Theoretical and Applied Electrodynamics, Russian Academy of Sciences, Moscow, 125412, Russia}
\author{O.V. Astafiev}%
\affiliation{Skolkovo Institute of Science and Technology, Nobel St. 3, 143026 Moscow, Russia}
\affiliation{Laboratory of Artificial Quantum Systems, \\ Moscow Institute of Physics and Technology, 141700 Dolgoprudny, Russia}
\date{\today}

\begin{abstract}
A cascade of two-level superconducting artificial atoms --- a source and a probe --- strongly coupled to a semi-infinite waveguide is a promising tool for observing nontrivial phenomena in quantum nonlinear optics. The probe atom can scatter an antibunched radiation emitted from the source, thereby generating a field with specific properties. We experimentally demonstrate wave mixing between nonclassical light from the coherently cw-pumped source and another coherent wave acting on the probe. We observe unique features in the wave mixing stationary spectrum which differs from mixing spectrum of two classical waves on the probe. These features are well described by adapting the theory \cite{Gardiner1994Driving} for a strongly coupled cascaded system of two atoms. We further analyze the theory to predict non-classical mixing spectra for various ratios of atoms' radiative constants. Both experimental and numerical results confirm the domination of multi-photon scattering process with only a single photon from the source. We evaluate entanglement of atoms in the quasistationary state and illustrate the connection between the expected second-order correlation function of the \textit{source} field and wave mixing side peaks corresponding to a certain number of scattered photons.

\end{abstract}

\maketitle

\section{Introduction}  
Applications of non-classical light \cite{mandel1986non-class,Strekalov2019Quantum} have led to remarkable progress in quantum optics and photonics over the past several decades. Quantum states of light are used in gravitational wave astronomy \cite{goda2008quantum}, quantum information processing \cite{ourjoumtsev2006generating}, and the quantum internet \cite{wehner2018quantum}. The reliable characterization of non-classical electromagnetic fields is necessary for many of these applications. Along this way, there are subtleties arising from the probabilistic nature of quantum theory \cite{holevo2011probabilistic}. An illustrative example is the measurement of a single propagating photon \cite{hadfield2009single}. Even if a perfect single-photon detector registers a photocount, it does not guarantee that a single-photon state was present prior to the measurement. This is due to the collapse of wave function \cite{Bassi2013Models}, which inevitably occurs during photodetection and irrevocably changes the state of the field.

To overcome these problems, methods for measuring time-dependent correlation functions \cite{glauber1963quantum,Grangier_1986} were developed. The practical approach to measuring a single photon is the use of the Hanbury-Brown-Twiss interferometer \cite{HBT}, which splits the field between two detectors. The corpuscular property does not allow the photon to be simultaneously registered in both channels, which leads to full anti-correlation of photocount sequences \cite{KimbleMandel}. In other words, the second-order correlation function of the output fields is zero. This is a common approach for ensuring that a single photon was present at the input, and a reliable way to characterize fields with non-classical photon statistics. This approach has conceptually evolved into a variety of tomographic methods \cite{lvovsky2009continuous}, allowing the restoration of phase-space distributions \cite{kirchmair2013observation, Bao2022experimental} of fields or full-photon statistics \cite{eichler2011experimental,houck2007generating}, relying either on photocounting detectors or on homodyne or heterodyne measurements of fields \cite{leonhardt1996sampling} in the time domain.

A single-photon signal of gigahertz frequency carries a small energy. Despite proof-of-concept implementations \cite{kono2018quantum, besse2018single}, practical photon detectors in this range are lacking. Measurement of the second-order correlation function of a linear field requires raising the field to the fourth power, worsening the signal-to-noise ratio. An alternative is to use the cross-Kerr interaction between the signal and trial modes \cite{PhysRevD.32.3208, PhysRevA.28.2646}. This non-destructive method detects the number of photons in the signal mode by measuring the amplitude of the coherent state in the probe mode. It avoids interference circuits with multiple detectors \cite{lang2013correlations} or extensive data processing \cite{lu2021quantum, zhou2020tunable}. Instead, it relies on four-wave mixing, requiring strong nonlinearity, which is difficult to achieve with modes of visible optical range \cite{PhysRevLett.113.173601}. These proposals \cite{PhysRevD.32.3208, PhysRevA.28.2646} have not been experimentally tested. However, such interaction can be implemented for propagating fields using a superconducting qubit strongly coupled with microwave modes in the waveguide \cite{Astafiev2010, Abdumalikov2011}. Several works \cite{PhysRevLett.111.053601, mirhosseini2019cavity} have shown the strong non-linearity of qubits in a waveguide to fields with single-photon amplitudes. Wave mixing with a qubit in the transmission line was demonstrated \cite{Dmitriev2017,Dmitriev2019,Vasenin2022}, and theoretical analysis \cite{Pogosov2021} revealed that coherent side peaks are connected with the photon statistics of the propagating fields. The method is called Quantum Wave Mixing (QWM) and it allows one to observe new features of light. For weak microwave signals, this method may be more applicable than measuring the correlation function or performing tomography.

In this work, we experimentally investigate QWM of a classical coherent wave and a quantum wave, both scattered on a single superconducting qubit. We construct a cascade system \cite{Vasenin2020, Gunin2023} where radiation from a strongly coupled source qubit is directed to a probe qubit on the same waveguide. A continuous pump with slight detuning is applied to the source, while the probe is driven by a signal with opposite detuning. We analyze the probe's emitted field spectra and observe significant deviations from the wave mixing spectra of two classical signals \cite{Dmitriev2019} on a single artificial atom. Using the master equation for the cascade atomic system \cite{kolobov1987quantum, gardiner1993driving}, we obtain numerical mixing spectra consistent with experimental observations. The model also examines regimes with different coupling constants of the source and probe. We infer that unique wave-mixing spectrum features could indicate photon antibunching from the source without directly measuring the second-order correlation function. Furthermore, we elucidate the relationship between $g^{(2)}(0)$ and coherent mixing peaks, discussing their interrelation.

Although the cascaded systems have already been tested experimentally \cite{pfaff2014unconditional,Meyer2015, Delteil2017}, it is important to note that the focus is often not on the optical effects in these systems, but on practical aspects relevant for the applications, such as the population transfer between atoms as nodes of common quantum network, or characterization of the net concurrence between atoms. Our goal is to study the system in the context of optics, that is, to measure the emission by the way which is most convenient for our platform and use this results for making conclusions about field properties, and more generally, about using suggested approach for the detection of non-classical light. 

The main part of the article is structured as follows. In section II we present the concept of our experiment. To do this, we reveal the analogy between our setup and a possible implementation in the atomic optics of visible range. Then in the same section we elucidate how the probe senses the photon statistics of field emitted by the source. Next, in section III we characterise the atomic cascade with low-power spectroscopic measurements. In section IV we avoid the source atom and present the results of wave mixing of two classical coherent tones on the probe, showing the agreement with the analytical model of this process. In section V we adopt a theoretical approach based on master equation which is being able to describe specific interaction of the probe with the field emitted by the source. In section VI we present the spectrum of the probe's emission  -- the result of wave mixing between classical tone and quantum emission of the source, both applied to the probe. We thoroughly compare these spectra to the ones presented in section III and conclude that in the cascaded case, multi-photon processes tend to involve only a single photon emitted by the source. Section VII describes the results of numerical evaluation of sideband peak intensities in the case of cascaded system, and there we focus on varying the coupling constants of the source and the probe. This analysis strongly supports our conclusion on the sensitivity of the probe to the photon statistics of the source's emission. Section VIII shows the results of perturbative calculations which generally confirm the correctness of numerical analysis. In conclusion, we summarize the results and discuss the perspectives of our approach for the nonlinear quantum optics of complex systems.
\begin{figure}[h]
\includegraphics[width=1\linewidth]{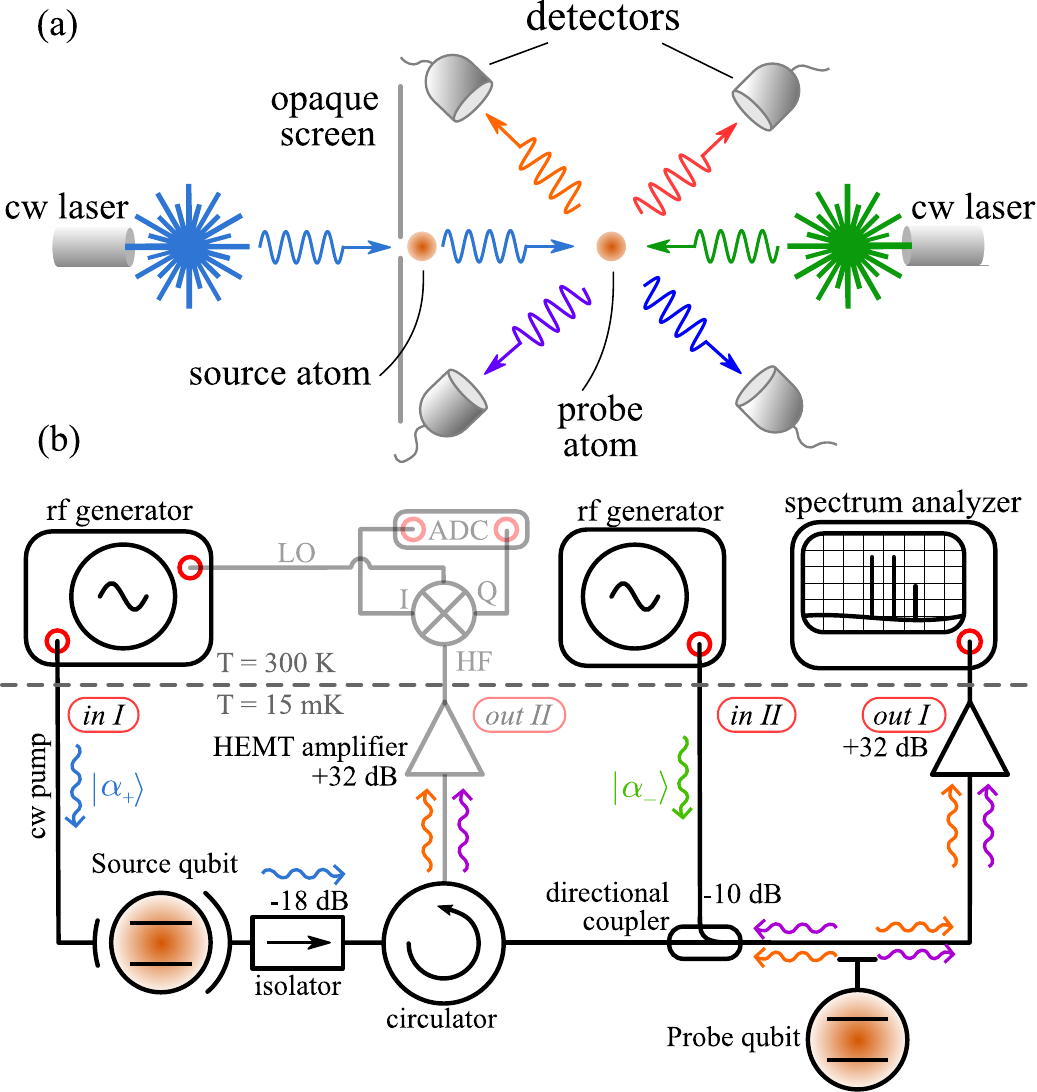}
\caption{\label{fig:phys-descr}(a) The experiment's optical concept: the probe atom scatters two coherent fields—non-classical from a source through a small aperture in an opaque screen, and classical from an external generator. The probe's field is detected and analyzed. (b) Simplified sketch of the waveguide-QED microwave setup with two superconducting transmon qubits in a dilution refrigerator. The measurements in this work are done with the use of  the channel labeled as ``out I''. }
\end{figure}

\section{The concept} 
In Fig.~\ref{fig:phys-descr}(a) we present a scheme of the cascaded system in paradigm of traditional atomic optics. The source atom is ``held'' in the open space and is effectively driven by a continuous-wave monochromatic laser light. In front of the source, there is an opaque screen with a sub-wavelength hole, so small that only rapidly vanishing evanescent waves can pass through the hole and excite the source effectively, that is, with the Rabi frequency much larger than the decay rate. We assume that most of the emission of the source could be redirected to another atom - the probe. We also assume that there is another cw laser that drives the probe atom without disturbing the source at all. Transition frequencies of both atoms and both cw lasers are close enough, that is, the detunings are much smaller than the natural linewidths of the atoms and Rabi frequencies of each atom driven by its own laser. This means that the probe is effectively driven by two fields: the coherent field of laser and the non-classical field from the source being in stationary state under the drive of another laser. Our intent is to study the coherent part of the field re-emitted by the probe on the condition that it is effectively detected either with linear or power detectors. Particularly, we are interested in demonstrating how non-classicality of source emission manifests itself in the spectrum of coherent emission of the probe. For comparison, another spectrum could be measured without opaque screen and without source atom, but with two slightly detuned laser beams scattered by a single probe atom. 

The presented scheme is difficult to construct with the use of real atoms: it is a problem to collect most of the source emission and scatter it on another atom. However, it is readily implemented in waveguide QED setup with a pair of superconducting qubits, Fig.~\ref{fig:phys-descr}(b). The source qubit is weakly coupled to a control line and strongly coupled to the emission line; both lines are semi-infinite coplanar waveguides on the chip. In other words, the capacitance between the control line and the source is relatively small, that free decay into this line due to voltage vacuum fluctuations is negligible, but as soon as a strong coherent field from an external generator is applied, it drives the qubit and it undergoes Rabi dynamics. At the same time, the capacitance with the emission line is large enough, and the qubit is coupled with quantum fluctuations and decays there with the emission of field \cite{peng2016tuneable,zhou2020tunable}. This configuration resembles an atom behind a small hole in opaque screen, as the source is effectively excited via the control line, but all the field goes into the emission line and no driving field from the control line passes through. Next, the propagating field in the emission waveguide transmits through the cryogenic isolator and circulator and scatters on the probe qubit (hereafter simply the probe). The probe, in turn, is located on the other chip and side-coupled to the transmission line. The directional coupler allows one to apply classical tone to pump the probe, and necessarily, an isolator and a circulator prevent this field from back-acting the state of the source. The probe scatters the field forward and backward along the waveguide, and we can collect what was transmitted or reflected from the probe and carefully analyze it. In this work, we mostly use the spectral analyzer, but digitizing after down-conversion is also available \cite{Vasenin2022}. To construct a cascade system, we utilize a pair of tunable transmon qubits with the sweetspot transition frequencies $\omega_{ge}/2\pi\approx 5.1$ GHz (slightly changing after subsequent cooldowns) between ground and first excited states $\ket{g}$ and $\ket{e}$ and anharmonicity of 350 MHz. The description of devices could be found elsewhere \cite{Vasenin2020, Vasenin2022, Gunin2023}.

\begin{figure}[h]
\includegraphics[width=1\linewidth]{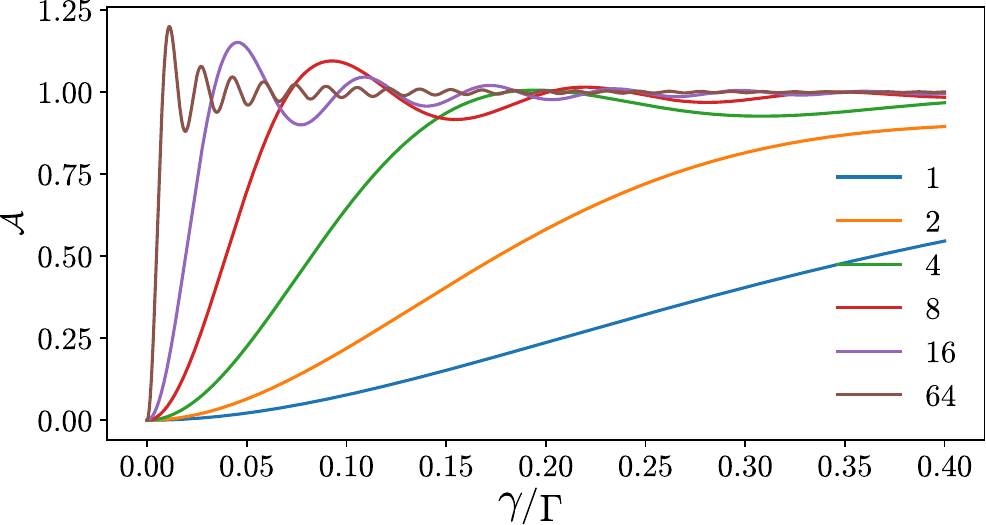}
\caption{\label{fig:antib} The effective antibunching $\mathcal{A}$ as function of coupling constant ratio $\gamma/\Gamma$ for various Rabi frequencies $\Omega/\gamma$ (see the legend) of the source's drive.    }
\end{figure}
In our scheme, the probe acts as the detector of a quantum field emitted by the source. Thereby it is instructive to analyze the photon-photon correlations in this quantum field in order to reveal its connection with wave mixing spectra, which will be demonstrated below. The signal emitted by a driven two-level system is known to be antibunched in the time domain \cite{PhysRevLett.125.170402, Kimble1977}, which is conventionally characterized with a second-order correlation function
\begin{equation}
g^{(2)}(\tau) = \frac{\braket{\sigma_+(0)\sigma_+(\tau)\sigma_-(\tau)\sigma_-{(0)}}}{((1+\braket{\sigma_z})/2)^2}.
\end{equation}
A solution of the optical Bloch equations gives the following expression for stationary emission: 
\begin{equation}
g^{(2)}(\tau) = 1-e^{-\frac{3}{4}\gamma\tau}\left[\cos(\Omega_g \tau)+\frac{3\gamma}{\Omega_g}\sin(\Omega_g \tau)\right], 
\end{equation}
where $\Omega_g=\sqrt{\Omega^2-\gamma^2/16}$ is generalized Rabi frequency, $\Omega=\mu V_0/\hbar$ is Rabi frequency of input propagating field with voltage amplitude $V_0$ resonant with $\ket{g}-\ket{e}$ transition,$\mu$ is the atomic dipole coupling moment to the input line, $\gamma$ is the radiative decay rate of the source to its emission line. Thus, $g^{(2)}(0)$ is always equal to $0$. For very large delays, $\tau \gg 1/\gamma$, we always get $g^{(2)}(\tau) \rightarrow 1$. However, in case $\Omega > \gamma/4$, $g^{(2)}$ as a function of $\tau$ oscillates between $0$ and $2$ then approaches $1$, and for small $\Omega$ it slowly and aperiodically rises from $0$ to $1$. In our setup, the signal emitted by the source scatters on the probe with natural radiative linewidth $\Gamma$, caused by two-sided decay to the same line, which is emission line for the source. To see how incoming field is perceived by our probe, we introduce the following parameter --- the degree of effective antibunching:
\begin{equation}
\mathcal{A} = \Gamma\int\limits_0^{1/\Gamma}g^{(2)}(\tau)d\tau,
\label{eq: antib}
\end{equation}
which characterizes how photon correlations in the field are persuaded by the probe, in other words, illustrates the degree of effective antibunching of source's light as seen by the probe. This feature relates $\mathcal{A}$ to commonly used Mandel parameter $Q$ \cite{Mandel:79,Treussart2002direct}, but for our setup, $\mathcal{A}$ depends on the probe natural linewidth.  Plots of $\mathcal{A}(1/\Gamma)$ according to Eq. \eqref{eq: antib}   are shown in Fig.~\ref{fig:antib}. If $\Gamma \ll \gamma$, then $\mathcal{A} \approx 1$ for any $\Omega$, but for $\Gamma \gg \gamma$ it also depends on $\Omega$. Particularly, it could be shown that for $\Omega \gg \gamma$ we get $\mathcal{A} \to 1$ in case $\Gamma \gg \Omega$, and $ 1.22 < \mathcal{A} < 0$ in case $\Gamma \sim \Omega$, whereas for $\Omega < \gamma$ it is always true that $0 < \mathcal{A} < 1$. 
Consequently, for small probe coupling, $\Gamma \ll \gamma$, the probe observes the classical statistics within the source's emission. For $\Gamma \gg \gamma, \Omega$, in most cases the probe detects antibunching, however, for specific ratios of $\Omega$ and $\Gamma$, the photon bunching might be observed. 
This arguments should be considered only as qualitative: we do not calculate explicitly the statistics of source photons absorbed by the probe. We now describe how our cascade system operates, and then show how the wave mixing visualizes the interaction between the probe and the field from the source.

\section{Low-power Spectroscopy} 
Our goal is to observe the mixing of coherent waves with a classical and a quantum signal. To achieve this, there are several prerequisites: (i) the source is efficiently excited via the control line; (ii) the source efficiently decays into the emission line; (iii) the signal from the source is transmitted to the probe with acceptable losses; and (iv) both the classical field applied to the probe and the field emitted by the probe do not disturb the source. The points (i)-(iii) are checked with transmission spectroscopy. To do this, we apply a weak coherent tone to input I. The tone is almost resonant to the source, whereas the probe is detuned. Measuring the frequency-dependent S-parameter from input I to output I, we observe the Lorentzian peak emitted by the source in the stationary state. Having that, we tune the probe qubit exactly in resonance with the source, and then the transmission dip owing to the reflection from the probe appears on top of the source's peak; see Fig.~\ref{fig:sts_cascade}. The measured transmission is expressed as:
\begin{figure}[t]
\includegraphics[width=1\linewidth]{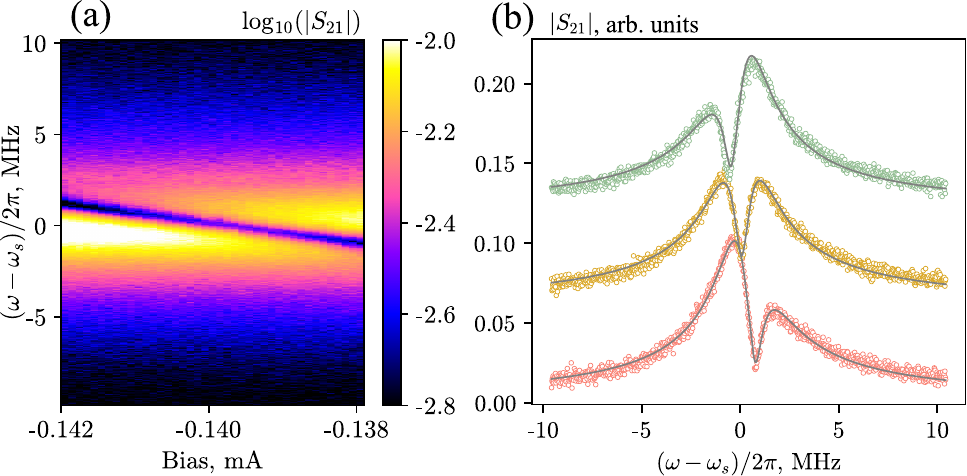}
\caption{\label{fig:sts_cascade}(a) The measured frequency-dependent transmission of low power wave applied to the input of the cascade system. The probe qubit is tuned to be in exact resonance with the source. (b) The transmission profiles for specific bias values fitted with the model. Fitting parameters are $\gamma/2\pi = 1.74,\: \gamma_\varphi/2\pi = 0.15,\: \Gamma/2\pi = 1.70,\: \Gamma_\varphi/2\pi = 0.19$~MHz, $ (\omega_p-\omega_s)/2\pi = [0.60, -0.01,-0.51] $, given for orange, yellow and green points, respectively. The vertical offset is added to each line for clarity. }
\end{figure}
\begin{equation}
t = \alpha \sqrt{A G}t_s t_p, 
\label{eq: trans}
\end{equation}
where $\alpha$ is amplitude losses between the source and the probe, $A$ is total attenuation on the way from signal output to the input capacitance of the source, $G$ is total amplification on the way from output of the probe to the signal input, $t_p$ and $t_s$ is the effective transmission coefficient of the probe and the source, respectively. Assuming weak drive of the source, we can write transmission as: 
\begin{eqnarray}
t_s = \frac{\gamma}{2\gamma_2}\frac{C_c}{C_e}\frac{1+i\Delta\omega_s/\gamma_2}{1+(\Delta\omega_s/\gamma_2)^2},  \\
t_p = 1-\frac{\Gamma}{2\Gamma_2}\frac{1+i\Delta\omega_p/\Gamma_2}{1+(\Delta\omega_p/\Gamma_2)^2}.
\end{eqnarray}
\begin{figure*}[t]
\includegraphics[width=1\linewidth]{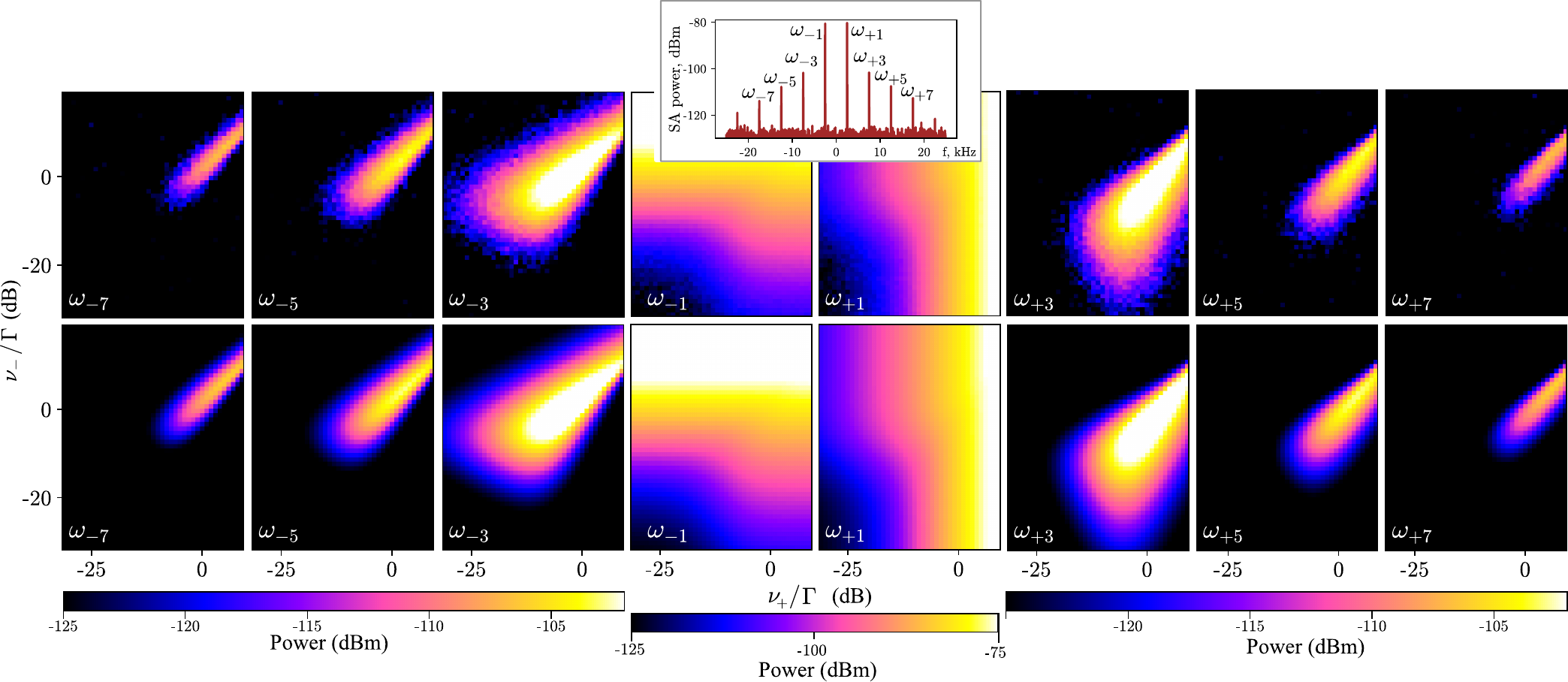}
\cprotect\caption{\label{fig:class_probe} The mixing of two classical tones of frequencies $\omega_+$ and $\omega_-$ on the probe. The tone of $\omega_+$ is applied through the input \texttt{in I} when the source is very detuned and its effect is negligible, see Fig.~\ref{fig:phys-descr}(b), and the tone of $\omega_-$ is applied through the input \texttt{in II}. The measured powers of the side peaks are plotted in the upper row of the panels as functions of $\nu_-/\Gamma$ and $\nu_+/\Gamma$. The inset panel in between components $\omega_{-1}$ and $\omega_{+1}$ of the upper row contains the single-run measured spectrum of mixing for $\nu_-/\Gamma = \nu_+/\Gamma = 0$~dB. The bottom panels depict analytical expressions of Eq.~\eqref{eq: intensities} recalculated to the power of the side peaks for $\Gamma=2$~MHz. All components are re-scaled by a single multiplier to fit the data.   }
\end{figure*}
Here, $C_c$ is the capacitance between the control waveguide and the source, $C_e$ is the capacitance between the source and output waveguide, $\gamma_2 = \gamma/2 +\gamma_\varphi$ and $\Gamma_2 = \Gamma/2 +\Gamma_\varphi$ are total dephasing rates, $\gamma_\varphi$ and $\Gamma_\varphi$ are pure dephasing constants for the source and the probe, respectively. In addition, we introduced detunings $\Delta\omega_{p,s} = \omega_{p,s} - \omega$, where $\omega$ is the frequency of the signal. We measure $t(\omega)$ dependence for a number of $\omega_p$ values for fixed $\omega_s$, see Fig. \ref{fig:sts_cascade}(a). Eq.~\eqref{eq: trans} allows fitting the data and extracting the qubit parameters, see Fig. \ref{fig:sts_cascade}(b). From the fit we extract $\Gamma$, $\gamma$ and also estimate $\Gamma_\varphi$ and $\gamma_\varphi$ of the probe and the source, respectively. Note that $t_s$ is not a real transmission of the coherent pump, because typically for our sources we have $C_c / C_e = 0.1$ and $C_e \approx 5 $~ fF, which makes leakage from weak pump negligible \cite{peng2016tuneable}. Only coherent emission of the source is measured at the output. The value of $\alpha\sqrt{AG}$ could be estimated in the following way. When the probe is detuned ($t_p=1$), we apply an input power of -45 dBm, which delivers a weak power on chip and does not saturate the source. We measure the maximal amplitude of source's emission line to be $|t|_{\text{max}} = \alpha\sqrt{AG}{C_c}/{C_e} = 0.013$, which gives $\alpha\sqrt{AG}=0.13$. With $A\approx -85$~dB for our input lines and $G\approx 75$ dB for used amplification cascade, we get $\alpha \approx 0.5$. This is a rough estimation and it could be improved with wave mixing measurements. The well-fitted double-resonance transmission confirms that the source and the probe does not disturb each other for small powers of the drive.

\section{Mixing of two classical waves}
To obtain a reference for comparison with quantum mixing, we proceed to the measurements of classical wave mixing on the probe. To do that, we detune the probe and send two coherent tones with Rabi amplitudes $\Omega_{\pm}$ and frequencies $\omega_{\pm} = \omega\pm \delta\omega$ to the input I , where $\delta\omega \ll \Gamma$ is a small detuning and we choose deliberately $\omega=\omega_p$. Many side peaks with frequencies $\omega_{\pm(2p+1)} = (p+1)\omega_{\pm} -p\omega_{\mp}$ appear as a result of scattering. The details of this experiment can be found in \cite{Dmitriev2017,Dmitriev2019}. We now measure the side-peak amplitudes as functions of the pump photon number per probe qubit's lifetime $\varkappa_{\pm}=\nu_{\pm}/\Gamma= \Omega_{\pm}^2/2\Gamma^2$, where $\nu_{\pm}=\Omega_{\pm}^2/2\Gamma$ is the photon flux. The results are given in Fig.~\ref{fig:class_probe} along with the exact analytical dependence for $\varkappa_\pm$ adopted from \cite{Dmitriev2019}:
\begin{multline}
\sqrt{\varkappa^{\:\text{sc}}_{\pm(2p+1)}} = \sqrt{\varkappa_\pm}\delta_{p0} + \\ +\frac{(-1)^p}{8\sqrt{\varkappa_-\varkappa_+}}\tan\theta\tan^p\frac{\theta}{2}\left(\sqrt{\varkappa_{\pm}}\tan\frac{\theta}{2}-\sqrt{\varkappa_{\mp}}\right),
\label{eq: intensities}
\end{multline} 
where $\theta = \arcsin\Big(\frac{4\sqrt{\varkappa_+\varkappa_-}}{\Gamma_2/\Gamma + 2(\varkappa_- + \varkappa_+)}\Big)$.
Good agreement is observed, and here we note the specific shape of this density plots: each of the peaks is maximal in a specific simply connected parameter region defined by approximate condition $\Omega_+\approx\Omega_-\approx p\Gamma $ and gradually decreases for larger Rabi amplitudes. Moreover, for any non-vanishing values of $\Omega_+$ and $\Omega_-$ (which means $\Omega_{\pm} \ge \Gamma$) all the peaks are present, although some orders may be very small due to $\Omega_+ \gg \Omega_-$ or vice versa. We stress that the analytical dependencies in Fig.~\ref{fig:class_probe} are derived under the assumption of a classical monochromatic pump. Overall, multi-photon scattering processes are becoming less probable with increasing the number of photons, and correspondingly, the peaks at frequencies with larger $p$ are generally less intensive than for smaller $p$. See also \cite{Dmitriev2019} for a detailed exploration of classical wave mixing.

\section{Theoretical analysis of cascaded system} 
\begin{figure}[h]
    \centering
    \includegraphics[width=0.83\linewidth]{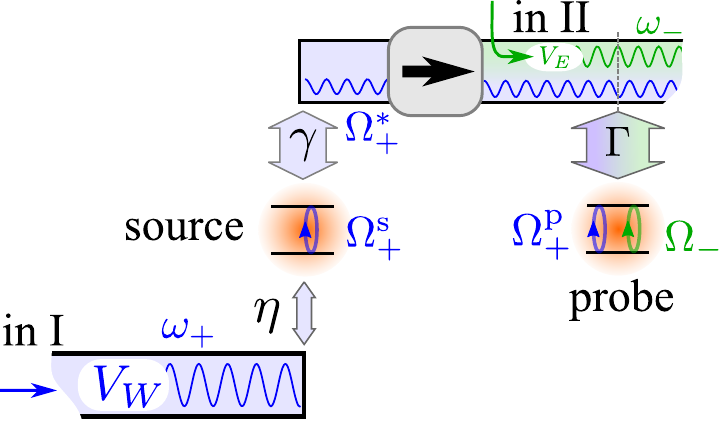}
    \caption{The schematic image of theoretically described two-atomic cascade. Semi-infinite waveguides are shown by half-cut rectangles, couplings are shown as two-sided arrows.  }
    \label{fig:scheme_theory}
\end{figure}
Before we proceed to QWM on the probe within the cascaded atomic setup, it is necessary to outline theoretical description for this kind of systems. We adopt the master equation approach \cite{kolobov1987quantum,gardiner1993driving,Gardiner1994Driving} for the conceptual scheme in Fig.~\ref{fig:scheme_theory}. The source is coupled to a pair of half-infinite waveguides: the left one is weakly coupled with rate $\eta$, and the right one is strongly coupled with $\gamma \gg \eta$. The probe is side-coupled with the second (right) waveguide with rate $\Gamma$. All decay rates $\gamma, \Gamma, \eta$ are due to corresponding dipole coupling moments denoted as $\mu_\gamma, \mu_\Gamma, \mu_\eta$. Input classical wave with a voltage $V_We^{i\omega_+ t}$ applied from \texttt{in I} drives the source with a Rabi amplitude $\Omega^\mathrm{s}_+ = \mu_{\eta}V_W/\hbar$. Under this drive, the source emit the nonclassical signal. The coherent part of it has an effective amplitude of $\Omega^*_+=-i\gamma\braket{\sigma^{-}_s(t)}$, see Fig.~\ref{fig:scheme_theory}. and there is also incoherent part, altogether making this wave non-classical. It then reaches the probe and also interacts with it. If it had been a classical wave (which is not our case, see also \cite{Gunin2023, Carreno2022Loss}), the probe would have been rotated by it with the Rabi amplitude $\Omega^\mathrm{p}_+={\mu_\Gamma}\Omega^*_+/{\mu_\gamma}$, see Fig.~\ref{fig:scheme_theory}. These two last statements give an oversimplified picture and are not to be applied straightforwardly, but we will use them below for the qualitative analysis of experimental and numerical results. And finally, there is another classical wave $V_Ee^{i\omega_-t}$ applied via \texttt{in II} and drives the probe with a Rabi amlpitude $\Omega_-$, providing the second component to generate wave mixing.

For the purposes of rigorous theoretical analysis, we denote the amplitudes of classical fields applied to waveguides in order to drive qubits as $W$ and $E$, respectively, and they are directly proportional to the voltage up to dimensional constant: $W = V_W/\sqrt{\hbar\omega Z_0}$, and the same stands for $E$.  The Hamiltonian of the problem is:
\begin{equation}
    \mc{\hat H}
    =
    \mc{\hat H_{\rm q}}
    +
    \mc{\hat H_{\rm dr,s}}
    +
    \mc{\hat H_{\rm dr,p}}
    ,
\end{equation}
where the Hamiltonian of non-interacting qubits is  
\begin{equation} \mc{\hat H_{\rm q}} =\frac{1}{2}
    \left(
        \omega_s \sigma^z_s
        +
        \omega_p \sigma^z_p
    \right).
\end{equation}
The driving part of the Hamiltonian is written as:
\begin{align}
    \mc{\hat H_{\rm dr,s}}
    = &\,
    -i \sqrt{\eta} \left(
        W \hat \sigma^+_s e^{i (\omega_d - \delta\omega)t}
        +
        \rm{c.c.}
    \right)
    ,
    \\
    \mc{\hat H_{\rm dr,p}}
    = &\,
    -i \sqrt{\frac{\Gamma}{2}} \left(
	    E \hat \sigma^+_p e^{i (\omega_d + \delta\omega) t}
	    +
	    \rm{c.c.}
    \right)
    .
\end{align}
Now we need to include not only the radiative relaxation of each atom, but also the fact that the probe is irradiated by all the field emitted by the source. As shown \cite{gardiner1993driving}, this task is accomplished by the standard form of the quantum master equation for the two-qubit density matrix $\rho$:
\begin{equation}
    \frac{\partial}{\partial t} \rho
    =
    -i \left[
        \mc{\hat H}, \rho
    \right]
    +
    \mc{\hat L} \rho
    .\label{eq:master}
\end{equation}
but with a very specific Lindblad term with a following form:
\begin{equation}
    \mc{\hat L} \rho
    =
    \mc{\hat L}_{s} \rho
    +
    \mc{\hat L}_{p} \rho
    +
    \mc{\hat L}_{sp} \rho
    .\label{eq: L_terms}
\end{equation}
The first two terms are straightforward and they describe the radiational relaxation of atoms because of waveguide modes:
\begin{multline}
    \mc{\hat L}_{s,p} \rho
    =
    \hat L_{s,p} \rho \hat L_{s,p}^{\dagger}
    -
    \frac{1}{2} \left\{
        \hat L_{s,p}^{\dagger} \hat L_{s,p}, \rho
    \right\}
    ,
    \\
    \hat L_{s} 
    = 
    \sqrt{\gamma + \eta} \sigma^-_s
    ,
    \hat L_{p} 
    = 
    \sqrt{\Gamma} \sigma^-_p
    ,\label{eq:L12}
\end{multline}
whereas the coupling term is non-hermitian and has the form:
\begin{equation}
    \mc{\hat L}_{sp} \rho
    = 
    \alpha\sqrt{\gamma \Gamma} \left(
        \left[
            \sigma_s^- \rho, \sigma^+_p
        \right]
        +
        \left[
             \sigma_p^-, \rho \sigma^+_s
        \right]
    \right)
    .\label{eq:L12int}
\end{equation}
Here, $\alpha$ is amplitude loss between the source and the probe. In absence of losses, $\alpha=1$. Here we note non-standard form of this term, comparing to the terms in Eq.~\eqref{eq:L12}. In spite of that, Eq.~\eqref{eq:L12int} preserves the hermitian property of the density matrix $\rho=\rho_s\otimes \rho_p$ of two qubits, because  $(\mc{\hat L}_{sp} \rho)^\dagger = \mc{\hat L}_{sp} \rho$. To clarify the meaning of that term, we provide a following argument. To account the action of the probe on the source (which is not the case in our setup), one need to include into Eq.~\eqref{eq: L_terms} the term $-\mc{\hat L}_{ps} \rho$, which is obtained from $\mc{\hat L}_{sp} \rho$ by a formal replacement $s \leftrightarrow p$. Then a sum of coupling terms $\mc{\hat L}_{sp} \rho - \mc{\hat L}_{ps} \rho$ will result in:
\begin{equation}
    \alpha\sqrt{\gamma \Gamma}\left[
        \hat{\sigma}_s^+ \hat{\sigma}_p^-
        -
        \hat{\sigma}_p^+ \hat{\sigma}_s^-
        , \hat \rho
    \right]
    = 
    -i \left[
        \mc{\hat H_{\rm int}}, \hat \rho   
    \right]
    , 
\end{equation}
where
\begin{equation}
    \mc{\hat H}_{\rm int} = -i  \alpha\sqrt{\gamma \Gamma} \left(
        \hat{\sigma}_p^+ \hat{\sigma}_s^-
        -
        \hat{\sigma}_s^+ \hat{\sigma}_p^-
    \right),
\end{equation}
that is, the hamiltonian of two dipole-coupled qubits. Therefore Eq.~\eqref{eq:L12int} accounts uni-directional coupling via the waveguide. One is able to show that the partial trace of Eq.~\eqref{eq:L12int} over the probe is  $\mathrm{Tr}_p[\mc{L}_{\rm sp} \hat \rho] \equiv 0$, which means that the described interaction has no effect on the state of the source. In this sense, it is more correct not to say that the source and the probe are interacting, but instead say that the probe ``adapts'' to the source, whereas the source ``knows nothing'' about the probe. The last statement is only correct in terms of partial density matrix of the source, but generally, under strong drive there will be a correlation (and consequently, non-zero concurrence) between qubits, as they constitute the two-particle quantum system.  
\begin{figure}[h]
\includegraphics[width=1\linewidth]{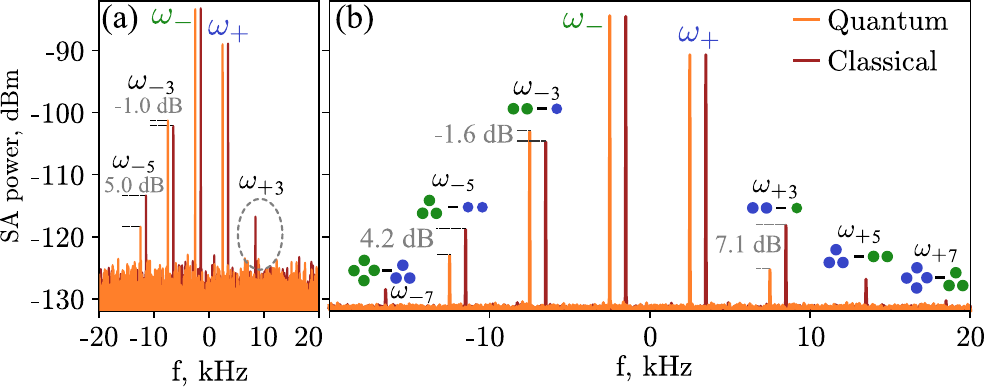}
\caption{\label{fig:q-cl spectrum} (a) The single-run measurement of coherent components for cascaded mixing at point $\nu_+/\gamma = -9.9$ dB and $\nu_-/\gamma = -4.0$ dB (orange trace) and for classical mixing (brown trace) for nearly same driving parameters, selected in a way that emitted components at $\omega_{\pm}$ are maximally equal. A small frequency offset is applied to the classical trace to make it visible. The difference in side peaks at $\omega_{-3}$ and $\omega_{-5}$ is labeled, and the absence of $\omega_{+3}$-component in cascaded mixing is highlighted with dashed oval. (b) The comparison of classical and quantum (cascaded) traces averaged over several values of driving strengths, around the parameters specified in (a). The classical trace is shifted by 1 kHz for clear comparison. The difference of corresponding components is labeled.   }
\end{figure}
\begin{figure*}[!t]
\includegraphics[width=1\linewidth]{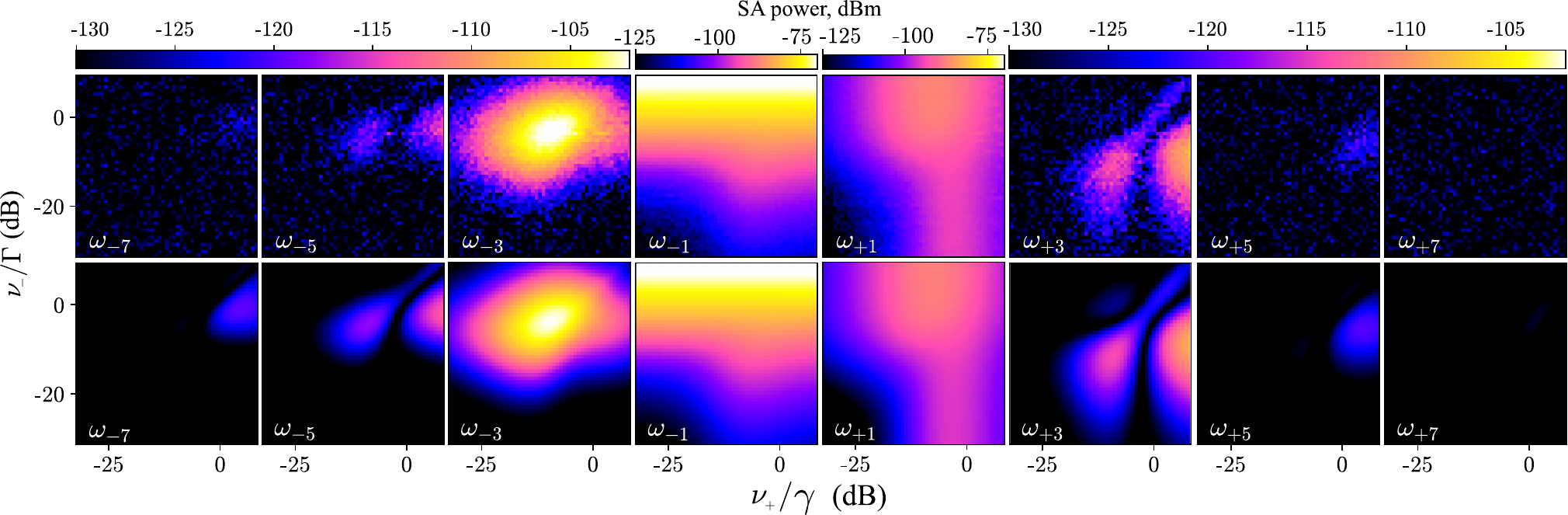}
\caption{\label{fig:cacs_mix} The side components emitted by the probe in wave mixing processes of stationary atomic emission at $\omega_+$ (emitted by the driven source) and classical tone at $\omega_-$, both applied to the probe. Upper row of panels: power of each side peak measured by spectral analyzer with a resolution bandwidth of 5 Hz as a function of $\nu_+/\gamma$ and $\nu_-/\Gamma$. Lower row of panels: the power fit obtained through numerical calculations. The fitting parameters are $\Gamma/2\pi=1.8 $~ MHz, $\gamma/2\pi=1.7$~ MHz, $\alpha=0.79$ and some general attenuation and amplification coefficients.   }
\end{figure*}

The numerical calculation of the coherent spectrum is done as follows. For any prescribed $E$ and $W$, we start from the ground state $\rho_0=\ket{0}\otimes\ket{0}$. Once the master equation is solved, we take the evolution of the system density operator $\rho(t)$ from $t=t_0$ to $t=t_0+nT$, where $t_0 \gg 1/\Gamma,1/\gamma$ is an arbitrary time and $T = 2\pi/\delta\omega$ is the period of precession of the quasi-stationary state of a two-qubit system. 
Using known $\rho(t)$, the output field of the source and probe is proportional to $\braket{\sigma^-_s}(t) = \mathrm{Tr}[\sigma^-_s \rho(t)]$ and $\braket{\sigma^-_p}(t) = \mathrm{Tr}[\sigma^-_p \rho(t)]$, respectively. This time-dependent quasi-stationary expectation value could easily be found numerically. The field that is eventually being detected is proportional to 
\begin{equation}
    \braket{a_\mathrm{out}}(t)=Ee^{i\omega_-t}+\sqrt\alpha{\gamma}\braket{\sigma^-_s}(t)+\sqrt{\frac{\Gamma}{2}}\braket{\sigma^-_p}(t).
\end{equation}
The squared Fourier spectrum of this output field gives us the desirable sideband peaks. 

As an exact analytical solution of the described master equation appears to be hardly tractable, we could build a perturbation theory. To do this, we neglect the correlations between qubits in the following way. As already explained, $\rho_s$ is not affected by the interaction term. Therefore, if we take $\mathrm{Tr}_p$ from Eq.~\eqref{eq:master}, what we get are standard Bloch equations for a $\rho_s$ of the source qubit driven by a slightly detuned field with constant amplitude. It has a known quasi-stationary solution which we denote as $\rho_s^{(0)}$. Then we use the following relation:
\begin{equation}
i \partial_t \hat \rho^{(n)}
    -
    \left[
    \mc{\hat H}, \hat \rho^{(n)}
    \right]
    -
    \left(\mc{L}_s + \mc{L}_p\right)\hat \rho ^{(n)}
    =
    \mc{L}_{ sp} \hat \rho ^{(n-1)}
    \label{eq: perturb_expr}
\end{equation}
to get the next order solution. Namely, for $n=1$ we take $\rho^{(0)} = \rho_s^{(0)} \otimes \rho_p^{(0)}$ and then take partial trace $\mathrm{Tr}_s$ of Eq.~\eqref{eq: perturb_expr} to get the set of equations for $\rho_p^{(1)}$. The solution for $\rho^{(1)}_p$ is then found with the use of Floquet representation. Assuming that $\rho_s^{(1)} \equiv \rho_s^{(0)}$, we finally find all frequency components of $\rho^{(1)} = \rho_s^{(0)} \otimes \rho_p^{(1)}$ and do the same to find higher order corrections. 

\section{Wave mixing of a classical wave with a quantum signal} 
Now we proceed to the wave mixing of classical and quantum signals. To do that, we tuned the qubits in resonance with each other and applied two coherent tones. One tone at $\omega_+$ is applied to the source via input I, resulting in continuous stationary emission coming from the source qubit, both coherent and incoherent. This emission then drives the probe qubit. The other tone at $\omega_-$ is applied directly to the probe via input II and summed with the source emission. In the spectra of the field re-emitted by the probe, we observe similar intermodulation peaks as for the probe driven by a bichromatic classical drive measured and described above in Fig.~\ref{fig:class_probe}. The single measurement of the spectrum is depicted as an orange trace in Fig.~\ref{fig:q-cl spectrum}(a). Note that this is the spectrum of wave mixing of classical and non-classical waves.

For this cascaded mixing (hereafter we refer to this case as the quantum case), the $\omega_+$-drive of the probe is, in fact, the emission of the source, and it is not arbitrarily controlled by changing the driving Rabi amplitude of source $\Omega^{\text{s}}_+$ as the source always emits a specific kind of signal. This signal is not classical; however, it has a coherent component with an equivalent of the classical amplitude given by \cite{Astafiev2010, peng2016tuneable}:
\begin{equation}
    \Omega^*_+ = -i \gamma \cdot \braket{\sigma_-},
\end{equation}
where $\braket{\sigma_-}$ is a function of $\gamma$ and $\Omega^{\text{s}}_+$. Therefore, to make thorough comparison, for classical set-up we adjust the amplitudes of probe drives $\Omega_+$ and $\Omega_-$ in order to provide the maximal coincidence between peaks at $\omega_-$ and $\omega_+$ for classical and quantum cases.  The results of the measurements are highlighted in Fig.~\ref{fig:q-cl spectrum}(a), where we measure two traces for fixed $\Omega_+$ (or $\Omega^{\text{s}}_+$) and $\Omega_-$, and also in Fig.~\ref{fig:q-cl spectrum}(b), where each trace is averaged over several values of two pumps, $\Omega_+$ and $\Omega_-$ for the classical case and $\Omega^\text{s}_+$ and $\Omega_-$ for the cascaded (quantum) case. Crudely speaking, we expect that the signals at $\omega_{\pm}$ in cascaded (or in opposite case, in classical) wave mixing are determined by a value of $\Omega^*_+, \Omega_-$ ($\Omega_{\pm}$) and by how much power is taken out of this components during multi-photon scattering processes on the probe. In this case, the side peaks raised in these processes will in fact be an indicator of how the signal emitted by the source differs from the classical wave.

We first analyze the $\omega_{-3}$ component. The emission in this component implies that the probe adsorbs two photons from its classical pump $\omega_-$, emits a single photon at $\omega_+$ as a stimulated emission affected by the signal from the source, and the sideband photon emerges at $\omega_{-3}=2\omega_--\omega_+$. For cascaded quantum mixing, the $\omega_{-3}$-component is larger than for classical wave mixing by 1.0 dB for a single trace and by 1.6 dB on average. In contrast, other components are significantly suppressed. The emission of $\omega_{-5}$ occurs when three photons from $\omega_-$ are absorbed and two of $\omega_+$ are emitted, and it is suppressed by 5.0 dB for a single trace (and 4.2 dB on average) compared to the classical one. Even more striking is that for $\omega_{+3}$, involving \textit{absorption} of two photons from the quantum signal at $\omega_+$, we also observe that in quantum case this component does not appear above the noise level for a single trace, and on average it is lower by 7.1 dB than its classical counterpart. The averaging of classical wave mixing spectrum allows us to see nonzero emission at $\omega_{-7}$, $\omega_{+5}$ and $\omega_{+7}$, however, they do not appear in the spectrum of cascaded mixing even after averaging; see Fig.~\ref{fig:q-cl spectrum}(b). These results tell us that the rate of two-photon absorption from source's stationary emission (that is, how often they involve multiphoton scattering on the probe) is much less than the same two-photon rate from the classical pump. This corresponds to the analysis based on Eq.~\eqref{eq: antib}: measurements in Fig.~\ref{fig:q-cl spectrum} are made in the regime $\Omega^{\rm{s}}_+ \le \gamma \approx \Gamma$, in which $\mathcal{A} \le 0.1$, hence the probe effectively observes strong antibunching, see Fig.~\ref{fig:antib}. Overall, this allows us to claim that by means of multiphoton processes we observe sub-Poissonian photon statistics of the source's emission.
\begin{figure*}[t]
	\includegraphics[width=1\linewidth]{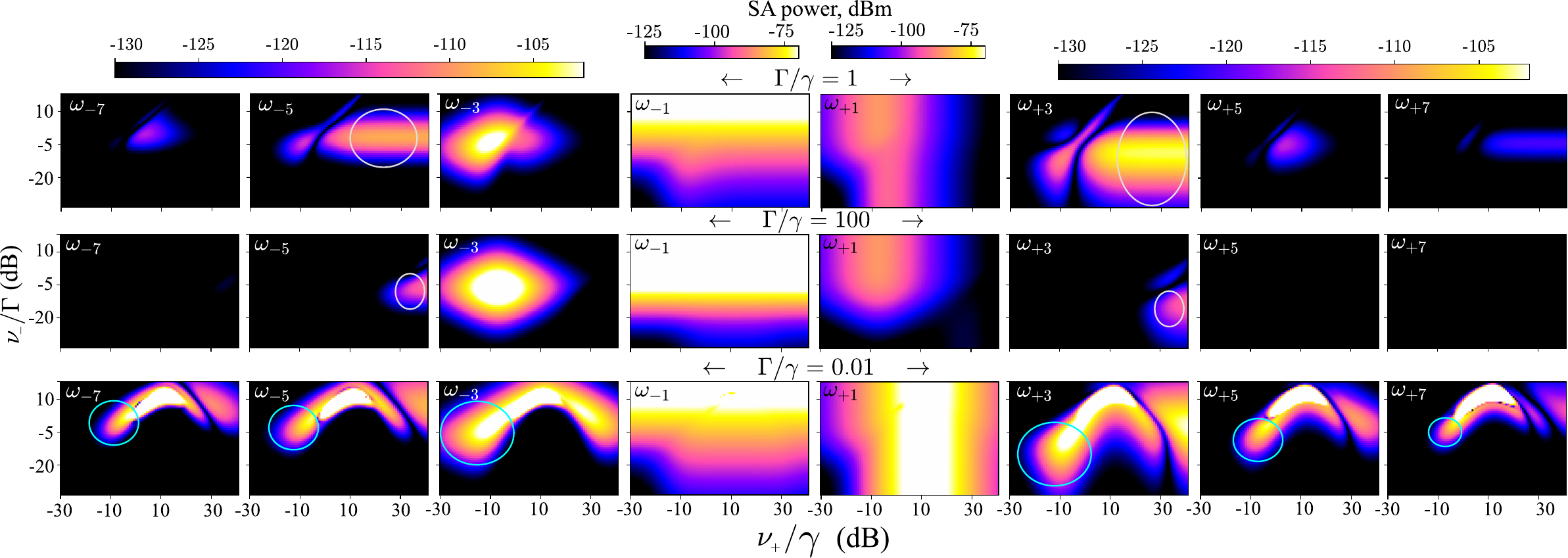}
	\caption{\label{fig:sim_all_regimes} The numerically simulated side peak intensities for ideal case of $\alpha=1$. Each row of panels represents various $\Gamma/\gamma$ ratios. For $\Gamma/\gamma = 1$, the absolute values of $\Gamma$ and $\gamma$ are taken to be the same as for simulations at Fig.~\ref{fig:cacs_mix}, that is: $\Gamma/2\pi=1.8 $~ MHz, $\gamma/2\pi=1.7$~ MHz. For other ratios, either $\Gamma$ or $\gamma$ is increased.  }
\end{figure*}

Next, we extend our analysis and measure the spectrum of coherent wave mixing in cascaded architecture over a wide range of pump amplitudes $\Omega^\text{s}_+$ and $\Omega_-$, similarly to how it was done for the classical case; see Fig.~\ref{fig:cacs_mix}. We observe that most of the sideband emission goes into the $\omega_{-3}$ component, which is bright across a wide range of amplitudes. Other sidebands are strongly suppressed: for $\omega_{+3}$ and $\omega_{-5}$, we notice a peculiar nonmonotonical behavior of sideband power, not inherent for classical mixing, where all dependencies have defined maximum. We detect negligible emission at $\omega_{+5}$ and $\omega_{\pm7}$, which is again in strong contrast to the classical case, where these peaks become relatively bright for strong drive. The bottom row of panels in Fig.~\ref{fig:cacs_mix} demonstrates a numerical solution of Eq.~\eqref{eq:master} for the optimal set of fitting parameters and for the corresponding range of pump rates (note that $W \propto \Omega_-$ and $E \propto \Omega^{\rm{s}}_+$). We see that the main features of the experiment are clearly reproduced with our model. We also note that the model appears to be extremely sensitive to $\alpha$, and we extract $\alpha=0.79$ from our fit, which correlates with our qualitative estimates of losses between qubits.

\section{Numerical simulations for various couplings}
The successful fitting confirms the applicability of the suggested theory for the analysis of cascaded wave mixing. Therefore, we can extend the results of numerical solution to analyze wave mixing for another ratio of radiative parameters of the qubits within the cascaded device, namely $\gamma$ and $\Gamma$. In our device, the coupling constants are fixed and could not be tuned \textit{in situ}. Using numerical modeling, we can examine extreme cases where $\gamma \gg \Gamma$ and $\gamma \ll \Gamma$. In principle, these cases also could be readily implemented in experiment with chips with different geometry, or even within the single device where couplings could be tuned via a magnetic field applied to dc-SQUID embedded in transmission lines or by any other possible approach. We present these results in Fig.~\ref{fig:sim_all_regimes}. We now discuss several observations from these simulations and describe and analyze some consequences that might be important for understanding the rich physics of the effect.

When $\gamma \gg \Gamma$, the source emission could be interpreted as a strong drive for the probe. Indeed, since the Rabi amplitude of this emission $\Omega_+^* \propto \gamma$, we could get \cite{Gunin2023} that the probe is effectively driven by this emission with an amplitude of $\Omega_+^p \propto \alpha\sqrt{\gamma\Gamma} \gg \Gamma$  -- which is to some extent analogous to a strong classical drive. Accordingly, we might expect a lot of sidebands to appear in the spectrum. However, the effective correlation time of the source emission is then much shorter than the response time of the probe $1/\Gamma$. Consequently, as the probe's response is averaged on its detection time, see Eq.~\eqref{eq: antib}, all non-classical features of source's emission, like antibunching, will be hardly detected by the probe. Therefore, we could expect that the power of sidebands for this case will be very close to classical picture; see Fig.~\ref{fig:class_probe}, but still not identical to it, because one of the drives is still not equivalent to classical tone. Indeed, in the bottom row of the panels of Fig.~\ref{fig:sim_all_regimes}, we clearly observe the regions denoted by blue ovals, where the sidebands behave exactly as for classical mixing. However, this takes place only for relatively low incoming power of both drives, and once the probe and, especially, the source is driven strongly, the picture of mixing gets more complicated.

\begin{figure}[h]
\includegraphics[width=1\linewidth]{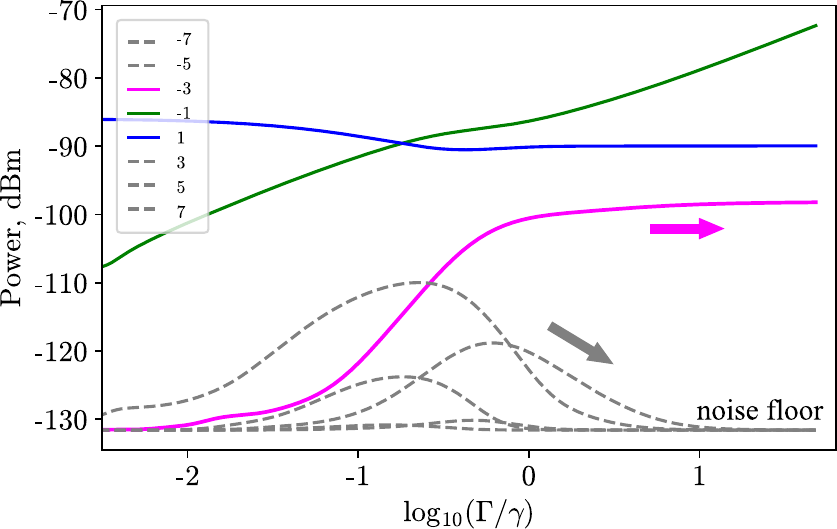}
\caption{\label{fig:sim_slice} Numerically simulated side peak intensities for the wave mixing in cascaded system. $\alpha=1$, $\nu_+/\gamma = \nu_-/\Gamma = -8 \text{ dB}$. The order $\pm(2p+1)$ of each component is labeled in the legend. For $\Gamma/\gamma \gg 1$ the nonzero emission (being well above noise floor of -131 dBm) is observed only at $\omega_{-3} = 2\omega_- - \omega_+$, which is the process (indicated by magenta arrow) with single-photon contribution from the source irradiated at $\omega_+$.}
\end{figure}
When $\Gamma \gg \gamma$, the probe becomes a truly broadband detector and is sensitive to anti-bunching of source emission. Indeed, in the middle row of panels in Fig.~\ref{fig:sim_all_regimes}, there is a domination of component $\omega_{-3}$, the hallmark of the four-photon process where a single photon from $\omega_{+}$ participates. We specifically illustrate this feature by increasing the ratio $\Gamma/\gamma$ while keeping the effective photon flux fixed, see Fig.~\ref{fig:sim_slice}. In fact, from Eq.~\eqref{eq: antib} we expect that $\mathcal{A}\approx 0$ provided that $\Gamma \gg \gamma$. However, for a very strong drive of the source, the detectable emission appears at $\omega_{-5}$ and $\omega_{+3}$ (the regions marked by gray ovals), implying that the processes with two or even four photons from $\omega_{+}$ also contribute to coherent scattering. Again, from Fig.~\ref{fig:antib} we conclude that the bunching could be observed in the case of very large $\Omega^{\rm{s}}_+$. Therefore, it confirms that the wave mixing on a single atom is capable of revealing the photon statistics of incoming fields. However, the variety of possible regimes illustrates the complexity of multiphoton scattering within a cascaded atomic system.

\section{Perturbative approach}
\begin{figure*}
\includegraphics[width=0.8\linewidth]{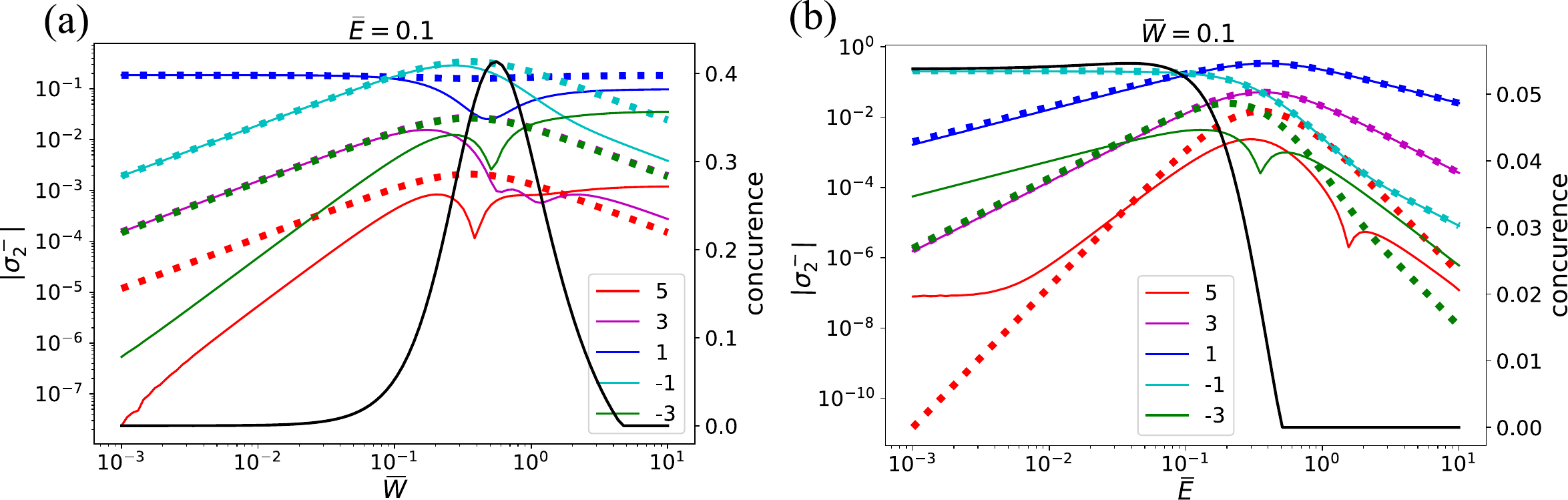}
\caption{\label{fig:conc-pert} The emission components of the probe depending (a) on driving amplitude $\bar{W}$ for fixed amplitude $\bar{E}$, and (b), vice versa. The parameters are: $\alpha=0.8, \gamma/2\pi=2, \Gamma/2\pi=4, \delta\omega/2\pi=0.01$. Dotted colored lines are the net emission in second order of perturbation theory, solid colored lines are exact numerical solutions. Black solid line is maximal absolute value of concurrence calculated from numerical solution.  }
\end{figure*}
As briefly described in the theory section, we build a perturbation theory to obtain analytical results for the side peak amplitudes and to compare them with numerical results. For simplicity, here we compare only the expressions for $\braket{\sigma^-_p}$, leaving out all prefactors. We are able to derive the expressions for the emission in zeroth, first and second orders of perturbation:
\begin{widetext}
\begin{align}
\braket{(\sigma^-_p)^{(0)}} &\approx \frac{2\bar{E}}{1+8 \bar{E} ^2} e^{-i\delta\omega t}, \quad 
\braket{(\sigma^-_p)^{(1)}} \approx \alpha\sqrt{\frac{\gamma}{\Gamma}}\frac{4\bar{W}}{(1+8\bar{E}^2)^2(1+8\bar{W}^2)} \left(-e^{i\delta\omega t} + 8\bar{E}^2e^{-3i\delta\omega t}\right),\\
\braket{(\sigma^-_p)^{(2)}} &\approx \alpha^2\frac{\gamma}{\Gamma}\frac{64\bar{W}}{(1+8\bar{E}^2)^4(1+8\bar{W}^2)}\left( e^{-i\delta\omega t} + e^{3i\delta\omega t} + 8\bar{E}^2 e^{-5i\delta\omega t}\right).
\end{align}
\end{widetext}

Here we introduced dimensionless amplitudes $\bar{W} = W\sqrt{\Gamma/2}$ and $\bar{E} = E\sqrt{\eta}$. Generally, we see that high-order components appear at high orders of perturbation series. In addition, we can compare the total emission obtained in numerical solutions, accounting for all the components, with the total emission obtained in perturbation theory. This result is outlined in Fig.~\ref{fig:conc-pert}, where the maximal absolute value of the concurrence of two qubits during the evolution cycle is also evaluated numerically. We see that the results of perturbation theory generally follow the exact numerical answer, but the coincidence takes place only for some components and only for specific driving regimes. Notably, as perturbative calculations neglect the qubit-qubit correlations, one might expect that a high concurrence between qubits in the stationary state indicates a significant discrepancy between exact amplitudes and perturbative ones. In fact, as seen in Fig.~\ref{fig:conc-pert}(a), we observe a large difference for the driving amplitudes $\bar{W}\approx 5\cdot10^{-1}$ and $\bar{E}=0.1$, where the concurrence approaches 0.4. However, the numerical simulations differ from the perturbation also for some parameters, where the concurrence is vanishingly small, see Fig.~\ref{fig:conc-pert}(b).

\section{Conclusion}
In conclusion, we have investigated the coherent intermodulation, or wave mixing, between classical tone and the non-classical output of two-level emitter, which is continuously driven by coherent wave. This interaction occurs with the use of another single two-level system - the probe -  embedded in the same waveguide. Our findings reveal that the spectrum diverges from that produced by two classical coherent tones, underscoring that processes involving a single photon from the quantum signal are more pronounced, whereas those involving two or more photons are suppressed. By employing the theory of cascaded quantum systems, we have provided a comprehensive description of our experimental setup. Furthermore, we utilized numerical solutions to explore intriguing scenarios of wave mixing in which the coupling constants vary by two orders of magnitude. Particularly, in the limit of large coupling of the probe, there are regime where all components are vanished except $\omega_{+3}$, indicating the probe's ability to detect the antibunching of incoming signal.  Our perturbative calculations exhibit partial agreement with the exact numerical results for emission side peaks. We believe that our results will significantly contribute to the rapidly developing field of nonlinear quantum optics with artificial atoms, particularly within superconducting quantum systems.

\begin{acknowledgments}
We wish to acknowledge the support of Russian Science Foundation: the development of the idea of experiment, design, fabrication, low temperature measurements, and data processing are sponsored by grant № 23–72–01052. W. V. P. acknowledges support from Grant of the Ministry of Science and Higher Education of the Russian Federation No. 075-15-2024-632 dated June 14, 2024 (development of perturbative and numerical calculations).  This work was performed using equipment from the MIPT Shared Facilities Center.
\end{acknowledgments}
\bibliography{bibliography}

\providecommand{\noopsort}[1]{}\providecommand{\singleletter}[1]{#1}%
\begin{thebibliography}{48}%
\makeatletter
\providecommand \@ifxundefined [1]{%
 \@ifx{#1\undefined}
}%
\providecommand \@ifnum [1]{%
 \ifnum #1\expandafter \@firstoftwo
 \else \expandafter \@secondoftwo
 \fi
}%
\providecommand \@ifx [1]{%
 \ifx #1\expandafter \@firstoftwo
 \else \expandafter \@secondoftwo
 \fi
}%
\providecommand \natexlab [1]{#1}%
\providecommand \enquote  [1]{``#1''}%
\providecommand \bibnamefont  [1]{#1}%
\providecommand \bibfnamefont [1]{#1}%
\providecommand \citenamefont [1]{#1}%
\providecommand \href@noop [0]{\@secondoftwo}%
\providecommand \href [0]{\begingroup \@sanitize@url \@href}%
\providecommand \@href[1]{\@@startlink{#1}\@@href}%
\providecommand \@@href[1]{\endgroup#1\@@endlink}%
\providecommand \@sanitize@url [0]{\catcode `\\12\catcode `\$12\catcode
  `\&12\catcode `\#12\catcode `\^12\catcode `\_12\catcode `\%12\relax}%
\providecommand \@@startlink[1]{}%
\providecommand \@@endlink[0]{}%
\providecommand \url  [0]{\begingroup\@sanitize@url \@url }%
\providecommand \@url [1]{\endgroup\@href {#1}{\urlprefix }}%
\providecommand \urlprefix  [0]{URL }%
\providecommand \Eprint [0]{\href }%
\providecommand \doibase [0]{https://doi.org/}%
\providecommand \selectlanguage [0]{\@gobble}%
\providecommand \bibinfo  [0]{\@secondoftwo}%
\providecommand \bibfield  [0]{\@secondoftwo}%
\providecommand \translation [1]{[#1]}%
\providecommand \BibitemOpen [0]{}%
\providecommand \bibitemStop [0]{}%
\providecommand \bibitemNoStop [0]{.\EOS\space}%
\providecommand \EOS [0]{\spacefactor3000\relax}%
\providecommand \BibitemShut  [1]{\csname bibitem#1\endcsname}%
\let\auto@bib@innerbib\@empty
\bibitem [{\citenamefont {Gardiner}\ and\ \citenamefont
  {Parkins}(1994)}]{Gardiner1994Driving}%
  \BibitemOpen
  \bibfield  {author} {\bibinfo {author} {\bibfnamefont {C.~W.}\ \bibnamefont
  {Gardiner}}\ and\ \bibinfo {author} {\bibfnamefont {A.~S.}\ \bibnamefont
  {Parkins}},\ }\href {https://doi.org/10.1103/PhysRevA.50.1792} {\bibfield
  {journal} {\bibinfo  {journal} {Phys. Rev. A}\ }\textbf {\bibinfo {volume}
  {50}},\ \bibinfo {pages} {1792} (\bibinfo {year} {1994})}\BibitemShut
  {NoStop}%
\bibitem [{\citenamefont {Mandel}(1986)}]{mandel1986non-class}%
  \BibitemOpen
  \bibfield  {author} {\bibinfo {author} {\bibfnamefont {L.}~\bibnamefont
  {Mandel}},\ }\href
  {https://iopscience.iop.org/article/10.1088/0031-8949/1986/T12/005}
  {\bibfield  {journal} {\bibinfo  {journal} {Physica Scripta}\ }\textbf
  {\bibinfo {volume} {1986}},\ \bibinfo {pages} {34} (\bibinfo {year}
  {1986})}\BibitemShut {NoStop}%
\bibitem [{\citenamefont {Strekalov}\ and\ \citenamefont
  {Leuchs}(2019)}]{Strekalov2019Quantum}%
  \BibitemOpen
  \bibfield  {author} {\bibinfo {author} {\bibfnamefont {D.~V.}\ \bibnamefont
  {Strekalov}}\ and\ \bibinfo {author} {\bibfnamefont {G.}~\bibnamefont
  {Leuchs}},\ }in\ \href {https://doi.org/10.1007/978-3-319-98402-5_3} {\emph
  {\bibinfo {booktitle} {Quantum Photonics: Pioneering Advances and Emerging
  Applications}}},\ \bibinfo {editor} {edited by\ \bibinfo {editor}
  {\bibfnamefont {R.~W.}\ \bibnamefont {Boyd}}, \bibinfo {editor}
  {\bibfnamefont {S.~G.}\ \bibnamefont {Lukishova}},\ and\ \bibinfo {editor}
  {\bibfnamefont {V.~N.}\ \bibnamefont {Zadkov}}}\ (\bibinfo  {publisher}
  {Springer International Publishing},\ \bibinfo {year} {2019})\ pp.\ \bibinfo
  {pages} {51--101}\BibitemShut {NoStop}%
\bibitem [{\citenamefont {Goda}\ \emph {et~al.}(2008)\citenamefont {Goda},
  \citenamefont {Miyakawa}, \citenamefont {Mikhailov}, \citenamefont {Saraf},
  \citenamefont {Adhikari}, \citenamefont {McKenzie}, \citenamefont {Ward},
  \citenamefont {Vass}, \citenamefont {Weinstein},\ and\ \citenamefont
  {Mavalvala}}]{goda2008quantum}%
  \BibitemOpen
  \bibfield  {author} {\bibinfo {author} {\bibfnamefont {K.}~\bibnamefont
  {Goda}}, \bibinfo {author} {\bibfnamefont {O.}~\bibnamefont {Miyakawa}},
  \bibinfo {author} {\bibfnamefont {E.~E.}\ \bibnamefont {Mikhailov}}, \bibinfo
  {author} {\bibfnamefont {S.}~\bibnamefont {Saraf}}, \bibinfo {author}
  {\bibfnamefont {R.}~\bibnamefont {Adhikari}}, \bibinfo {author}
  {\bibfnamefont {K.}~\bibnamefont {McKenzie}}, \bibinfo {author}
  {\bibfnamefont {R.}~\bibnamefont {Ward}}, \bibinfo {author} {\bibfnamefont
  {S.}~\bibnamefont {Vass}}, \bibinfo {author} {\bibfnamefont {A.~J.}\
  \bibnamefont {Weinstein}},\ and\ \bibinfo {author} {\bibfnamefont
  {N.}~\bibnamefont {Mavalvala}},\ }\href
  {https://www.nature.com/articles/nphys920} {\bibfield  {journal} {\bibinfo
  {journal} {Nature Physics}\ }\textbf {\bibinfo {volume} {4}},\ \bibinfo
  {pages} {472} (\bibinfo {year} {2008})}\BibitemShut {NoStop}%
\bibitem [{\citenamefont {Ourjoumtsev}\ \emph {et~al.}(2006)\citenamefont
  {Ourjoumtsev}, \citenamefont {Tualle-Brouri}, \citenamefont {Laurat},\ and\
  \citenamefont {Grangier}}]{ourjoumtsev2006generating}%
  \BibitemOpen
  \bibfield  {author} {\bibinfo {author} {\bibfnamefont {A.}~\bibnamefont
  {Ourjoumtsev}}, \bibinfo {author} {\bibfnamefont {R.}~\bibnamefont
  {Tualle-Brouri}}, \bibinfo {author} {\bibfnamefont {J.}~\bibnamefont
  {Laurat}},\ and\ \bibinfo {author} {\bibfnamefont {P.}~\bibnamefont
  {Grangier}},\ }\href {https://www.science.org/doi/10.1126/science.1122858}
  {\bibfield  {journal} {\bibinfo  {journal} {Science}\ }\textbf {\bibinfo
  {volume} {312}},\ \bibinfo {pages} {83} (\bibinfo {year} {2006})}\BibitemShut
  {NoStop}%
\bibitem [{\citenamefont {Wehner}\ \emph {et~al.}(2018)\citenamefont {Wehner},
  \citenamefont {Elkouss},\ and\ \citenamefont {Hanson}}]{wehner2018quantum}%
  \BibitemOpen
  \bibfield  {author} {\bibinfo {author} {\bibfnamefont {S.}~\bibnamefont
  {Wehner}}, \bibinfo {author} {\bibfnamefont {D.}~\bibnamefont {Elkouss}},\
  and\ \bibinfo {author} {\bibfnamefont {R.}~\bibnamefont {Hanson}},\ }\href
  {https://www.science.org/doi/full/10.1126/science.aam9288} {\bibfield
  {journal} {\bibinfo  {journal} {Science}\ }\textbf {\bibinfo {volume}
  {362}},\ \bibinfo {pages} {eaam9288} (\bibinfo {year} {2018})}\BibitemShut
  {NoStop}%
\bibitem [{\citenamefont {Holevo}(2011)}]{holevo2011probabilistic}%
  \BibitemOpen
  \bibfield  {author} {\bibinfo {author} {\bibfnamefont {A.~S.}\ \bibnamefont
  {Holevo}},\ }\href {https://link.springer.com/book/10.1007/978-88-7642-378-9}
  {\emph {\bibinfo {title} {Probabilistic and statistical aspects of quantum
  theory}}},\ Vol.~\bibinfo {volume} {1}\ (\bibinfo  {publisher} {Springer
  Science \& Business Media},\ \bibinfo {year} {2011})\BibitemShut {NoStop}%
\bibitem [{\citenamefont {Hadfield}(2009)}]{hadfield2009single}%
  \BibitemOpen
  \bibfield  {author} {\bibinfo {author} {\bibfnamefont {R.~H.}\ \bibnamefont
  {Hadfield}},\ }\href {https://www.nature.com/articles/nphoton.2009.230}
  {\bibfield  {journal} {\bibinfo  {journal} {Nature Photonics}\ }\textbf
  {\bibinfo {volume} {3}},\ \bibinfo {pages} {696} (\bibinfo {year}
  {2009})}\BibitemShut {NoStop}%
\bibitem [{\citenamefont {Bassi}\ \emph {et~al.}(2013)\citenamefont {Bassi},
  \citenamefont {Lochan}, \citenamefont {Satin}, \citenamefont {Singh},\ and\
  \citenamefont {Ulbricht}}]{Bassi2013Models}%
  \BibitemOpen
  \bibfield  {author} {\bibinfo {author} {\bibfnamefont {A.}~\bibnamefont
  {Bassi}}, \bibinfo {author} {\bibfnamefont {K.}~\bibnamefont {Lochan}},
  \bibinfo {author} {\bibfnamefont {S.}~\bibnamefont {Satin}}, \bibinfo
  {author} {\bibfnamefont {T.~P.}\ \bibnamefont {Singh}},\ and\ \bibinfo
  {author} {\bibfnamefont {H.}~\bibnamefont {Ulbricht}},\ }\href
  {https://link.aps.org/doi/10.1103/RevModPhys.85.471} {\bibfield  {journal}
  {\bibinfo  {journal} {Rev. Mod. Phys.}\ }\textbf {\bibinfo {volume} {85}},\
  \bibinfo {pages} {471} (\bibinfo {year} {2013})}\BibitemShut {NoStop}%
\bibitem [{\citenamefont {Glauber}(1963)}]{glauber1963quantum}%
  \BibitemOpen
  \bibfield  {author} {\bibinfo {author} {\bibfnamefont {R.~J.}\ \bibnamefont
  {Glauber}},\ }\href
  {https://journals.aps.org/pr/abstract/10.1103/PhysRev.130.2529} {\bibfield
  {journal} {\bibinfo  {journal} {Physical Review}\ }\textbf {\bibinfo {volume}
  {130}},\ \bibinfo {pages} {2529} (\bibinfo {year} {1963})}\BibitemShut
  {NoStop}%
\bibitem [{\citenamefont {Grangier}\ \emph {et~al.}(1986)\citenamefont
  {Grangier}, \citenamefont {Roger},\ and\ \citenamefont
  {Aspect}}]{Grangier_1986}%
  \BibitemOpen
  \bibfield  {author} {\bibinfo {author} {\bibfnamefont {P.}~\bibnamefont
  {Grangier}}, \bibinfo {author} {\bibfnamefont {G.}~\bibnamefont {Roger}},\
  and\ \bibinfo {author} {\bibfnamefont {A.}~\bibnamefont {Aspect}},\ }\href
  {https://doi.org/10.1209/0295-5075/1/4/004} {\bibfield  {journal} {\bibinfo
  {journal} {Europhysics Letters}\ }\textbf {\bibinfo {volume} {1}},\ \bibinfo
  {pages} {173} (\bibinfo {year} {1986})}\BibitemShut {NoStop}%
\bibitem [{\citenamefont {Brown}\ and\ \citenamefont {Twiss}(1956)}]{HBT}%
  \BibitemOpen
  \bibfield  {author} {\bibinfo {author} {\bibfnamefont {R.~H.}\ \bibnamefont
  {Brown}}\ and\ \bibinfo {author} {\bibfnamefont {R.~Q.}\ \bibnamefont
  {Twiss}},\ }\href {https://doi.org/10.1038/177027a0} {\bibfield  {journal}
  {\bibinfo  {journal} {Nature}\ }\textbf {\bibinfo {volume} {177}},\ \bibinfo
  {pages} {27} (\bibinfo {year} {1956})}\BibitemShut {NoStop}%
\bibitem [{\citenamefont {Kimble}\ \emph
  {et~al.}(1977{\natexlab{a}})\citenamefont {Kimble}, \citenamefont
  {Dagenais},\ and\ \citenamefont {Mandel}}]{KimbleMandel}%
  \BibitemOpen
  \bibfield  {author} {\bibinfo {author} {\bibfnamefont {H.~J.}\ \bibnamefont
  {Kimble}}, \bibinfo {author} {\bibfnamefont {M.}~\bibnamefont {Dagenais}},\
  and\ \bibinfo {author} {\bibfnamefont {L.}~\bibnamefont {Mandel}},\ }\href
  {https://doi.org/10.1103/PhysRevLett.39.691} {\bibfield  {journal} {\bibinfo
  {journal} {Phys. Rev. Lett.}\ }\textbf {\bibinfo {volume} {39}},\ \bibinfo
  {pages} {691} (\bibinfo {year} {1977}{\natexlab{a}})}\BibitemShut {NoStop}%
\bibitem [{\citenamefont {Lvovsky}\ and\ \citenamefont
  {Raymer}(2009)}]{lvovsky2009continuous}%
  \BibitemOpen
  \bibfield  {author} {\bibinfo {author} {\bibfnamefont {A.~I.}\ \bibnamefont
  {Lvovsky}}\ and\ \bibinfo {author} {\bibfnamefont {M.~G.}\ \bibnamefont
  {Raymer}},\ }\href
  {https://journals.aps.org/rmp/abstract/10.1103/RevModPhys.81.299} {\bibfield
  {journal} {\bibinfo  {journal} {Reviews of Modern Physics}\ }\textbf
  {\bibinfo {volume} {81}},\ \bibinfo {pages} {299} (\bibinfo {year}
  {2009})}\BibitemShut {NoStop}%
\bibitem [{\citenamefont {Kirchmair}\ \emph {et~al.}(2013)\citenamefont
  {Kirchmair}, \citenamefont {Vlastakis}, \citenamefont {Leghtas},
  \citenamefont {Nigg}, \citenamefont {Paik}, \citenamefont {Ginossar},
  \citenamefont {Mirrahimi}, \citenamefont {Frunzio}, \citenamefont {Girvin},\
  and\ \citenamefont {Schoelkopf}}]{kirchmair2013observation}%
  \BibitemOpen
  \bibfield  {author} {\bibinfo {author} {\bibfnamefont {G.}~\bibnamefont
  {Kirchmair}}, \bibinfo {author} {\bibfnamefont {B.}~\bibnamefont
  {Vlastakis}}, \bibinfo {author} {\bibfnamefont {Z.}~\bibnamefont {Leghtas}},
  \bibinfo {author} {\bibfnamefont {S.~E.}\ \bibnamefont {Nigg}}, \bibinfo
  {author} {\bibfnamefont {H.}~\bibnamefont {Paik}}, \bibinfo {author}
  {\bibfnamefont {E.}~\bibnamefont {Ginossar}}, \bibinfo {author}
  {\bibfnamefont {M.}~\bibnamefont {Mirrahimi}}, \bibinfo {author}
  {\bibfnamefont {L.}~\bibnamefont {Frunzio}}, \bibinfo {author} {\bibfnamefont
  {S.~M.}\ \bibnamefont {Girvin}},\ and\ \bibinfo {author} {\bibfnamefont
  {R.~J.}\ \bibnamefont {Schoelkopf}},\ }\href
  {https://www.nature.com/articles/nature11902} {\bibfield  {journal} {\bibinfo
   {journal} {Nature}\ }\textbf {\bibinfo {volume} {495}},\ \bibinfo {pages}
  {205} (\bibinfo {year} {2013})}\BibitemShut {NoStop}%
\bibitem [{\citenamefont {Bao}\ \emph {et~al.}(2022)\citenamefont {Bao},
  \citenamefont {Wang}, \citenamefont {Wu}, \citenamefont {Li}, \citenamefont
  {Cai}, \citenamefont {Wang}, \citenamefont {Ma}, \citenamefont {Cai},
  \citenamefont {Han}, \citenamefont {Wang}, \citenamefont {Song},
  \citenamefont {Sun}, \citenamefont {Zhang},\ and\ \citenamefont
  {Duan}}]{Bao2022experimental}%
  \BibitemOpen
  \bibfield  {author} {\bibinfo {author} {\bibfnamefont {Z.}~\bibnamefont
  {Bao}}, \bibinfo {author} {\bibfnamefont {Z.}~\bibnamefont {Wang}}, \bibinfo
  {author} {\bibfnamefont {Y.}~\bibnamefont {Wu}}, \bibinfo {author}
  {\bibfnamefont {Y.}~\bibnamefont {Li}}, \bibinfo {author} {\bibfnamefont
  {W.}~\bibnamefont {Cai}}, \bibinfo {author} {\bibfnamefont {W.}~\bibnamefont
  {Wang}}, \bibinfo {author} {\bibfnamefont {Y.}~\bibnamefont {Ma}}, \bibinfo
  {author} {\bibfnamefont {T.}~\bibnamefont {Cai}}, \bibinfo {author}
  {\bibfnamefont {X.}~\bibnamefont {Han}}, \bibinfo {author} {\bibfnamefont
  {J.}~\bibnamefont {Wang}}, \bibinfo {author} {\bibfnamefont {Y.}~\bibnamefont
  {Song}}, \bibinfo {author} {\bibfnamefont {L.}~\bibnamefont {Sun}}, \bibinfo
  {author} {\bibfnamefont {H.}~\bibnamefont {Zhang}},\ and\ \bibinfo {author}
  {\bibfnamefont {L.}~\bibnamefont {Duan}},\ }\href
  {https://doi.org/10.1103/PhysRevA.105.063717} {\bibfield  {journal} {\bibinfo
   {journal} {Phys. Rev. A}\ }\textbf {\bibinfo {volume} {105}},\ \bibinfo
  {pages} {063717} (\bibinfo {year} {2022})}\BibitemShut {NoStop}%
\bibitem [{\citenamefont {Eichler}\ \emph {et~al.}(2011)\citenamefont
  {Eichler}, \citenamefont {Bozyigit}, \citenamefont {Lang}, \citenamefont
  {Steffen}, \citenamefont {Fink},\ and\ \citenamefont
  {Wallraff}}]{eichler2011experimental}%
  \BibitemOpen
  \bibfield  {author} {\bibinfo {author} {\bibfnamefont {C.}~\bibnamefont
  {Eichler}}, \bibinfo {author} {\bibfnamefont {D.}~\bibnamefont {Bozyigit}},
  \bibinfo {author} {\bibfnamefont {C.}~\bibnamefont {Lang}}, \bibinfo {author}
  {\bibfnamefont {L.}~\bibnamefont {Steffen}}, \bibinfo {author} {\bibfnamefont
  {J.}~\bibnamefont {Fink}},\ and\ \bibinfo {author} {\bibfnamefont
  {A.}~\bibnamefont {Wallraff}},\ }\href
  {https://journals.aps.org/prl/abstract/10.1103/PhysRevLett.106.220503}
  {\bibfield  {journal} {\bibinfo  {journal} {Physical Review Letters}\
  }\textbf {\bibinfo {volume} {106}},\ \bibinfo {pages} {220503} (\bibinfo
  {year} {2011})}\BibitemShut {NoStop}%
\bibitem [{\citenamefont {Houck}\ \emph {et~al.}(2007)\citenamefont {Houck},
  \citenamefont {Schuster}, \citenamefont {Gambetta}, \citenamefont {Schreier},
  \citenamefont {Johnson}, \citenamefont {Chow}, \citenamefont {Frunzio},
  \citenamefont {Majer}, \citenamefont {Devoret}, \citenamefont {Girvin} \emph
  {et~al.}}]{houck2007generating}%
  \BibitemOpen
  \bibfield  {author} {\bibinfo {author} {\bibfnamefont {A.~A.}\ \bibnamefont
  {Houck}}, \bibinfo {author} {\bibfnamefont {D.}~\bibnamefont {Schuster}},
  \bibinfo {author} {\bibfnamefont {J.}~\bibnamefont {Gambetta}}, \bibinfo
  {author} {\bibfnamefont {J.}~\bibnamefont {Schreier}}, \bibinfo {author}
  {\bibfnamefont {B.}~\bibnamefont {Johnson}}, \bibinfo {author} {\bibfnamefont
  {J.}~\bibnamefont {Chow}}, \bibinfo {author} {\bibfnamefont {L.}~\bibnamefont
  {Frunzio}}, \bibinfo {author} {\bibfnamefont {J.}~\bibnamefont {Majer}},
  \bibinfo {author} {\bibfnamefont {M.}~\bibnamefont {Devoret}}, \bibinfo
  {author} {\bibfnamefont {S.}~\bibnamefont {Girvin}}, \emph {et~al.},\ }\href
  {https://www.nature.com/articles/nature06126} {\bibfield  {journal} {\bibinfo
   {journal} {Nature}\ }\textbf {\bibinfo {volume} {449}},\ \bibinfo {pages}
  {328} (\bibinfo {year} {2007})}\BibitemShut {NoStop}%
\bibitem [{\citenamefont {Leonhardt}\ \emph {et~al.}(1996)\citenamefont
  {Leonhardt}, \citenamefont {Munroe}, \citenamefont {Kiss}, \citenamefont
  {Richter},\ and\ \citenamefont {Raymer}}]{leonhardt1996sampling}%
  \BibitemOpen
  \bibfield  {author} {\bibinfo {author} {\bibfnamefont {U.}~\bibnamefont
  {Leonhardt}}, \bibinfo {author} {\bibfnamefont {M.}~\bibnamefont {Munroe}},
  \bibinfo {author} {\bibfnamefont {T.}~\bibnamefont {Kiss}}, \bibinfo {author}
  {\bibfnamefont {T.}~\bibnamefont {Richter}},\ and\ \bibinfo {author}
  {\bibfnamefont {M.}~\bibnamefont {Raymer}},\ }\href
  {https://www.sciencedirect.com/science/article/pii/0030401896000612}
  {\bibfield  {journal} {\bibinfo  {journal} {Optics Communications}\ }\textbf
  {\bibinfo {volume} {127}},\ \bibinfo {pages} {144} (\bibinfo {year}
  {1996})}\BibitemShut {NoStop}%
\bibitem [{\citenamefont {Kono}\ \emph {et~al.}(2018)\citenamefont {Kono},
  \citenamefont {Koshino}, \citenamefont {Tabuchi}, \citenamefont {Noguchi},\
  and\ \citenamefont {Nakamura}}]{kono2018quantum}%
  \BibitemOpen
  \bibfield  {author} {\bibinfo {author} {\bibfnamefont {S.}~\bibnamefont
  {Kono}}, \bibinfo {author} {\bibfnamefont {K.}~\bibnamefont {Koshino}},
  \bibinfo {author} {\bibfnamefont {Y.}~\bibnamefont {Tabuchi}}, \bibinfo
  {author} {\bibfnamefont {A.}~\bibnamefont {Noguchi}},\ and\ \bibinfo {author}
  {\bibfnamefont {Y.}~\bibnamefont {Nakamura}},\ }\href
  {https://www.nature.com/articles/s41567-018-0066-3} {\bibfield  {journal}
  {\bibinfo  {journal} {Nature Physics}\ }\textbf {\bibinfo {volume} {14}},\
  \bibinfo {pages} {546} (\bibinfo {year} {2018})}\BibitemShut {NoStop}%
\bibitem [{\citenamefont {Besse}\ \emph {et~al.}(2018)\citenamefont {Besse},
  \citenamefont {Gasparinetti}, \citenamefont {Collodo}, \citenamefont
  {Walter}, \citenamefont {Kurpiers}, \citenamefont {Pechal}, \citenamefont
  {Eichler},\ and\ \citenamefont {Wallraff}}]{besse2018single}%
  \BibitemOpen
  \bibfield  {author} {\bibinfo {author} {\bibfnamefont {J.-C.}\ \bibnamefont
  {Besse}}, \bibinfo {author} {\bibfnamefont {S.}~\bibnamefont {Gasparinetti}},
  \bibinfo {author} {\bibfnamefont {M.~C.}\ \bibnamefont {Collodo}}, \bibinfo
  {author} {\bibfnamefont {T.}~\bibnamefont {Walter}}, \bibinfo {author}
  {\bibfnamefont {P.}~\bibnamefont {Kurpiers}}, \bibinfo {author}
  {\bibfnamefont {M.}~\bibnamefont {Pechal}}, \bibinfo {author} {\bibfnamefont
  {C.}~\bibnamefont {Eichler}},\ and\ \bibinfo {author} {\bibfnamefont
  {A.}~\bibnamefont {Wallraff}},\ }\href
  {https://journals.aps.org/prx/abstract/10.1103/PhysRevX.8.021003} {\bibfield
  {journal} {\bibinfo  {journal} {Physical Review X}\ }\textbf {\bibinfo
  {volume} {8}},\ \bibinfo {pages} {021003} (\bibinfo {year}
  {2018})}\BibitemShut {NoStop}%
\bibitem [{\citenamefont {Walls}\ \emph {et~al.}(1985)\citenamefont {Walls},
  \citenamefont {Collet},\ and\ \citenamefont {Milburn}}]{PhysRevD.32.3208}%
  \BibitemOpen
  \bibfield  {author} {\bibinfo {author} {\bibfnamefont {D.~F.}\ \bibnamefont
  {Walls}}, \bibinfo {author} {\bibfnamefont {M.~J.}\ \bibnamefont {Collet}},\
  and\ \bibinfo {author} {\bibfnamefont {G.~J.}\ \bibnamefont {Milburn}},\
  }\href {https://doi.org/10.1103/PhysRevD.32.3208} {\bibfield  {journal}
  {\bibinfo  {journal} {Phys. Rev. D}\ }\textbf {\bibinfo {volume} {32}},\
  \bibinfo {pages} {3208} (\bibinfo {year} {1985})}\BibitemShut {NoStop}%
\bibitem [{\citenamefont {Milburn}\ and\ \citenamefont
  {Walls}(1983)}]{PhysRevA.28.2646}%
  \BibitemOpen
  \bibfield  {author} {\bibinfo {author} {\bibfnamefont {G.~J.}\ \bibnamefont
  {Milburn}}\ and\ \bibinfo {author} {\bibfnamefont {D.~F.}\ \bibnamefont
  {Walls}},\ }\href {https://doi.org/10.1103/PhysRevA.28.2646} {\bibfield
  {journal} {\bibinfo  {journal} {Phys. Rev. A}\ }\textbf {\bibinfo {volume}
  {28}},\ \bibinfo {pages} {2646} (\bibinfo {year} {1983})}\BibitemShut
  {NoStop}%
\bibitem [{\citenamefont {Lang}\ \emph {et~al.}(2013)\citenamefont {Lang},
  \citenamefont {Eichler}, \citenamefont {Steffen}, \citenamefont {Fink},
  \citenamefont {Woolley}, \citenamefont {Blais},\ and\ \citenamefont
  {Wallraff}}]{lang2013correlations}%
  \BibitemOpen
  \bibfield  {author} {\bibinfo {author} {\bibfnamefont {C.}~\bibnamefont
  {Lang}}, \bibinfo {author} {\bibfnamefont {C.}~\bibnamefont {Eichler}},
  \bibinfo {author} {\bibfnamefont {L.}~\bibnamefont {Steffen}}, \bibinfo
  {author} {\bibfnamefont {J.}~\bibnamefont {Fink}}, \bibinfo {author}
  {\bibfnamefont {M.~J.}\ \bibnamefont {Woolley}}, \bibinfo {author}
  {\bibfnamefont {A.}~\bibnamefont {Blais}},\ and\ \bibinfo {author}
  {\bibfnamefont {A.}~\bibnamefont {Wallraff}},\ }\href
  {https://www.nature.com/articles/nphys2612} {\bibfield  {journal} {\bibinfo
  {journal} {Nature Physics}\ }\textbf {\bibinfo {volume} {9}},\ \bibinfo
  {pages} {345} (\bibinfo {year} {2013})}\BibitemShut {NoStop}%
\bibitem [{\citenamefont {Lu}\ \emph {et~al.}(2021)\citenamefont {Lu},
  \citenamefont {Bengtsson}, \citenamefont {Burnett}, \citenamefont {Suri},
  \citenamefont {Sathyamoorthy}, \citenamefont {Nilsson}, \citenamefont
  {Scigliuzzo}, \citenamefont {Bylander}, \citenamefont {Johansson},\ and\
  \citenamefont {Delsing}}]{lu2021quantum}%
  \BibitemOpen
  \bibfield  {author} {\bibinfo {author} {\bibfnamefont {Y.}~\bibnamefont
  {Lu}}, \bibinfo {author} {\bibfnamefont {A.}~\bibnamefont {Bengtsson}},
  \bibinfo {author} {\bibfnamefont {J.~J.}\ \bibnamefont {Burnett}}, \bibinfo
  {author} {\bibfnamefont {B.}~\bibnamefont {Suri}}, \bibinfo {author}
  {\bibfnamefont {S.~R.}\ \bibnamefont {Sathyamoorthy}}, \bibinfo {author}
  {\bibfnamefont {H.~R.}\ \bibnamefont {Nilsson}}, \bibinfo {author}
  {\bibfnamefont {M.}~\bibnamefont {Scigliuzzo}}, \bibinfo {author}
  {\bibfnamefont {J.}~\bibnamefont {Bylander}}, \bibinfo {author}
  {\bibfnamefont {G.}~\bibnamefont {Johansson}},\ and\ \bibinfo {author}
  {\bibfnamefont {P.}~\bibnamefont {Delsing}},\ }\href
  {https://www.nature.com/articles/s41534-021-00480-5} {\bibfield  {journal}
  {\bibinfo  {journal} {npj Quantum Information}\ }\textbf {\bibinfo {volume}
  {7}},\ \bibinfo {pages} {140} (\bibinfo {year} {2021})}\BibitemShut {NoStop}%
\bibitem [{\citenamefont {Zhou}\ \emph {et~al.}(2020)\citenamefont {Zhou},
  \citenamefont {Peng}, \citenamefont {Horiuchi}, \citenamefont {Astafiev},\
  and\ \citenamefont {Tsai}}]{zhou2020tunable}%
  \BibitemOpen
  \bibfield  {author} {\bibinfo {author} {\bibfnamefont {Y.}~\bibnamefont
  {Zhou}}, \bibinfo {author} {\bibfnamefont {Z.}~\bibnamefont {Peng}}, \bibinfo
  {author} {\bibfnamefont {Y.}~\bibnamefont {Horiuchi}}, \bibinfo {author}
  {\bibfnamefont {O.}~\bibnamefont {Astafiev}},\ and\ \bibinfo {author}
  {\bibfnamefont {J.}~\bibnamefont {Tsai}},\ }\href
  {https://www.nature.com/articles/ncomms12588} {\bibfield  {journal} {\bibinfo
   {journal} {Physical Review Applied}\ }\textbf {\bibinfo {volume} {13}},\
  \bibinfo {pages} {034007} (\bibinfo {year} {2020})}\BibitemShut {NoStop}%
\bibitem [{\citenamefont {Guerreiro}\ \emph {et~al.}(2014)\citenamefont
  {Guerreiro}, \citenamefont {Martin}, \citenamefont {Sanguinetti},
  \citenamefont {Pelc}, \citenamefont {Langrock}, \citenamefont {Fejer},
  \citenamefont {Gisin}, \citenamefont {Zbinden}, \citenamefont {Sangouard},\
  and\ \citenamefont {Thew}}]{PhysRevLett.113.173601}%
  \BibitemOpen
  \bibfield  {author} {\bibinfo {author} {\bibfnamefont {T.}~\bibnamefont
  {Guerreiro}}, \bibinfo {author} {\bibfnamefont {A.}~\bibnamefont {Martin}},
  \bibinfo {author} {\bibfnamefont {B.}~\bibnamefont {Sanguinetti}}, \bibinfo
  {author} {\bibfnamefont {J.~S.}\ \bibnamefont {Pelc}}, \bibinfo {author}
  {\bibfnamefont {C.}~\bibnamefont {Langrock}}, \bibinfo {author}
  {\bibfnamefont {M.~M.}\ \bibnamefont {Fejer}}, \bibinfo {author}
  {\bibfnamefont {N.}~\bibnamefont {Gisin}}, \bibinfo {author} {\bibfnamefont
  {H.}~\bibnamefont {Zbinden}}, \bibinfo {author} {\bibfnamefont
  {N.}~\bibnamefont {Sangouard}},\ and\ \bibinfo {author} {\bibfnamefont
  {R.~T.}\ \bibnamefont {Thew}},\ }\href
  {https://doi.org/10.1103/PhysRevLett.113.173601} {\bibfield  {journal}
  {\bibinfo  {journal} {Phys. Rev. Lett.}\ }\textbf {\bibinfo {volume} {113}},\
  \bibinfo {pages} {173601} (\bibinfo {year} {2014})}\BibitemShut {NoStop}%
\bibitem [{\citenamefont {Astafiev}\ \emph {et~al.}(2010)\citenamefont
  {Astafiev}, \citenamefont {Zagoskin}, \citenamefont {Abdumalikov},
  \citenamefont {Pashkin}, \citenamefont {Yamamoto}, \citenamefont {Inomata},
  \citenamefont {Nakamura},\ and\ \citenamefont {Tsai}}]{Astafiev2010}%
  \BibitemOpen
  \bibfield  {author} {\bibinfo {author} {\bibfnamefont {O.}~\bibnamefont
  {Astafiev}}, \bibinfo {author} {\bibfnamefont {A.~M.}\ \bibnamefont
  {Zagoskin}}, \bibinfo {author} {\bibfnamefont {A.~A.}\ \bibnamefont
  {Abdumalikov}}, \bibinfo {author} {\bibfnamefont {Y.~A.}\ \bibnamefont
  {Pashkin}}, \bibinfo {author} {\bibfnamefont {T.}~\bibnamefont {Yamamoto}},
  \bibinfo {author} {\bibfnamefont {K.}~\bibnamefont {Inomata}}, \bibinfo
  {author} {\bibfnamefont {Y.}~\bibnamefont {Nakamura}},\ and\ \bibinfo
  {author} {\bibfnamefont {J.~S.}\ \bibnamefont {Tsai}},\ }\href
  {https://doi.org/10.1126/science.1181918} {\bibfield  {journal} {\bibinfo
  {journal} {Science}\ }\textbf {\bibinfo {volume} {327}},\ \bibinfo {pages}
  {840} (\bibinfo {year} {2010})}\BibitemShut {NoStop}%
\bibitem [{\citenamefont {Abdumalikov}\ \emph {et~al.}(2011)\citenamefont
  {Abdumalikov}, \citenamefont {Astafiev}, \citenamefont {Pashkin},
  \citenamefont {Nakamura},\ and\ \citenamefont {Tsai}}]{Abdumalikov2011}%
  \BibitemOpen
  \bibfield  {author} {\bibinfo {author} {\bibfnamefont {A.~A.}\ \bibnamefont
  {Abdumalikov}}, \bibinfo {author} {\bibfnamefont {O.~V.}\ \bibnamefont
  {Astafiev}}, \bibinfo {author} {\bibfnamefont {Y.~A.}\ \bibnamefont
  {Pashkin}}, \bibinfo {author} {\bibfnamefont {Y.}~\bibnamefont {Nakamura}},\
  and\ \bibinfo {author} {\bibfnamefont {J.~S.}\ \bibnamefont {Tsai}},\ }\href
  {https://doi.org/10.1103/PhysRevLett.107.043604} {\bibfield  {journal}
  {\bibinfo  {journal} {Phys. Rev. Lett.}\ }\textbf {\bibinfo {volume} {107}},\
  \bibinfo {pages} {043604} (\bibinfo {year} {2011})}\BibitemShut {NoStop}%
\bibitem [{\citenamefont {Hoi}\ \emph {et~al.}(2013)\citenamefont {Hoi},
  \citenamefont {Kockum}, \citenamefont {Palomaki}, \citenamefont {Stace},
  \citenamefont {Fan}, \citenamefont {Tornberg}, \citenamefont {Sathyamoorthy},
  \citenamefont {Johansson}, \citenamefont {Delsing},\ and\ \citenamefont
  {Wilson}}]{PhysRevLett.111.053601}%
  \BibitemOpen
  \bibfield  {author} {\bibinfo {author} {\bibfnamefont {I.-C.}\ \bibnamefont
  {Hoi}}, \bibinfo {author} {\bibfnamefont {A.~F.}\ \bibnamefont {Kockum}},
  \bibinfo {author} {\bibfnamefont {T.}~\bibnamefont {Palomaki}}, \bibinfo
  {author} {\bibfnamefont {T.~M.}\ \bibnamefont {Stace}}, \bibinfo {author}
  {\bibfnamefont {B.}~\bibnamefont {Fan}}, \bibinfo {author} {\bibfnamefont
  {L.}~\bibnamefont {Tornberg}}, \bibinfo {author} {\bibfnamefont {S.~R.}\
  \bibnamefont {Sathyamoorthy}}, \bibinfo {author} {\bibfnamefont
  {G.}~\bibnamefont {Johansson}}, \bibinfo {author} {\bibfnamefont
  {P.}~\bibnamefont {Delsing}},\ and\ \bibinfo {author} {\bibfnamefont {C.~M.}\
  \bibnamefont {Wilson}},\ }\href
  {https://doi.org/10.1103/PhysRevLett.111.053601} {\bibfield  {journal}
  {\bibinfo  {journal} {Phys. Rev. Lett.}\ }\textbf {\bibinfo {volume} {111}},\
  \bibinfo {pages} {053601} (\bibinfo {year} {2013})}\BibitemShut {NoStop}%
\bibitem [{\citenamefont {Mirhosseini}\ \emph {et~al.}(2019)\citenamefont
  {Mirhosseini}, \citenamefont {Kim}, \citenamefont {Zhang}, \citenamefont
  {Sipahigil}, \citenamefont {Dieterle}, \citenamefont {Keller}, \citenamefont
  {Asenjo-Garcia}, \citenamefont {Chang},\ and\ \citenamefont
  {Painter}}]{mirhosseini2019cavity}%
  \BibitemOpen
  \bibfield  {author} {\bibinfo {author} {\bibfnamefont {M.}~\bibnamefont
  {Mirhosseini}}, \bibinfo {author} {\bibfnamefont {E.}~\bibnamefont {Kim}},
  \bibinfo {author} {\bibfnamefont {X.}~\bibnamefont {Zhang}}, \bibinfo
  {author} {\bibfnamefont {A.}~\bibnamefont {Sipahigil}}, \bibinfo {author}
  {\bibfnamefont {P.~B.}\ \bibnamefont {Dieterle}}, \bibinfo {author}
  {\bibfnamefont {A.~J.}\ \bibnamefont {Keller}}, \bibinfo {author}
  {\bibfnamefont {A.}~\bibnamefont {Asenjo-Garcia}}, \bibinfo {author}
  {\bibfnamefont {D.~E.}\ \bibnamefont {Chang}},\ and\ \bibinfo {author}
  {\bibfnamefont {O.}~\bibnamefont {Painter}},\ }\href
  {https://www.nature.com/articles/s41586-019-1196-1} {\bibfield  {journal}
  {\bibinfo  {journal} {Nature}\ }\textbf {\bibinfo {volume} {569}},\ \bibinfo
  {pages} {692} (\bibinfo {year} {2019})}\BibitemShut {NoStop}%
\bibitem [{\citenamefont {Dmitriev}\ \emph {et~al.}(2017)\citenamefont
  {Dmitriev}, \citenamefont {Shaikhaidarov}, \citenamefont {Antonov},
  \citenamefont {H{\"{o}}nigl-Decrinis},\ and\ \citenamefont
  {Astafiev}}]{Dmitriev2017}%
  \BibitemOpen
  \bibfield  {author} {\bibinfo {author} {\bibfnamefont {A.~Y.}\ \bibnamefont
  {Dmitriev}}, \bibinfo {author} {\bibfnamefont {R.}~\bibnamefont
  {Shaikhaidarov}}, \bibinfo {author} {\bibfnamefont {V.~N.}\ \bibnamefont
  {Antonov}}, \bibinfo {author} {\bibfnamefont {T.}~\bibnamefont
  {H{\"{o}}nigl-Decrinis}},\ and\ \bibinfo {author} {\bibfnamefont {O.~V.}\
  \bibnamefont {Astafiev}},\ }\href
  {https://doi.org/10.1038/s41467-017-01471-x} {\bibfield  {journal} {\bibinfo
  {journal} {Nature Communications}\ }\textbf {\bibinfo {volume} {8}},\
  \bibinfo {pages} {1} (\bibinfo {year} {2017})}\BibitemShut {NoStop}%
\bibitem [{\citenamefont {Dmitriev}\ \emph {et~al.}(2019)\citenamefont
  {Dmitriev}, \citenamefont {Shaikhaidarov}, \citenamefont
  {H{\"{o}}nigl-Decrinis}, \citenamefont {{De Graaf}}, \citenamefont
  {Antonov},\ and\ \citenamefont {Astafiev}}]{Dmitriev2019}%
  \BibitemOpen
  \bibfield  {author} {\bibinfo {author} {\bibfnamefont {A.~Y.}\ \bibnamefont
  {Dmitriev}}, \bibinfo {author} {\bibfnamefont {R.}~\bibnamefont
  {Shaikhaidarov}}, \bibinfo {author} {\bibfnamefont {T.}~\bibnamefont
  {H{\"{o}}nigl-Decrinis}}, \bibinfo {author} {\bibfnamefont {S.~E.}\
  \bibnamefont {{De Graaf}}}, \bibinfo {author} {\bibfnamefont {V.~N.}\
  \bibnamefont {Antonov}},\ and\ \bibinfo {author} {\bibfnamefont {O.~V.}\
  \bibnamefont {Astafiev}},\ }\href
  {https://doi.org/10.1103/PhysRevA.100.013808} {\bibfield  {journal} {\bibinfo
   {journal} {Phys. Rev. A}\ }\textbf {\bibinfo {volume} {100}},\ \bibinfo
  {pages} {1} (\bibinfo {year} {2019})}\BibitemShut {NoStop}%
\bibitem [{\citenamefont {Vasenin}\ \emph {et~al.}(2022)\citenamefont
  {Vasenin}, \citenamefont {Dmitriev}, \citenamefont {Kadyrmetov},
  \citenamefont {Bolgar},\ and\ \citenamefont {Astafiev}}]{Vasenin2022}%
  \BibitemOpen
  \bibfield  {author} {\bibinfo {author} {\bibfnamefont {A.~V.}\ \bibnamefont
  {Vasenin}}, \bibinfo {author} {\bibfnamefont {A.~Y.}\ \bibnamefont
  {Dmitriev}}, \bibinfo {author} {\bibfnamefont {S.~V.}\ \bibnamefont
  {Kadyrmetov}}, \bibinfo {author} {\bibfnamefont {A.~N.}\ \bibnamefont
  {Bolgar}},\ and\ \bibinfo {author} {\bibfnamefont {O.~V.}\ \bibnamefont
  {Astafiev}},\ }\href {https://doi.org/10.1103/PhysRevA.106.L041701}
  {\bibfield  {journal} {\bibinfo  {journal} {Phys. Rev. A}\ }\textbf {\bibinfo
  {volume} {106}},\ \bibinfo {pages} {L041701} (\bibinfo {year}
  {2022})}\BibitemShut {NoStop}%
\bibitem [{\citenamefont {Pogosov}\ \emph {et~al.}(2021)\citenamefont
  {Pogosov}, \citenamefont {Dmitriev},\ and\ \citenamefont
  {Astafiev}}]{Pogosov2021}%
  \BibitemOpen
  \bibfield  {author} {\bibinfo {author} {\bibfnamefont {W.~V.}\ \bibnamefont
  {Pogosov}}, \bibinfo {author} {\bibfnamefont {A.~Y.}\ \bibnamefont
  {Dmitriev}},\ and\ \bibinfo {author} {\bibfnamefont {O.~V.}\ \bibnamefont
  {Astafiev}},\ }\href {https://doi.org/10.1103/PhysRevA.104.023703} {\bibfield
   {journal} {\bibinfo  {journal} {Phys. Rev. A}\ }\textbf {\bibinfo {volume}
  {104}},\ \bibinfo {pages} {023703} (\bibinfo {year} {2021})}\BibitemShut
  {NoStop}%
\bibitem [{\citenamefont {Vasenin}\ \emph {et~al.}(2020)\citenamefont
  {Vasenin}, \citenamefont {Dmitriev}, \citenamefont {Kadyrmetov},\ and\
  \citenamefont {Astafiev}}]{Vasenin2020}%
  \BibitemOpen
  \bibfield  {author} {\bibinfo {author} {\bibfnamefont {A.}~\bibnamefont
  {Vasenin}}, \bibinfo {author} {\bibfnamefont {A.}~\bibnamefont {Dmitriev}},
  \bibinfo {author} {\bibfnamefont {S.}~\bibnamefont {Kadyrmetov}},\ and\
  \bibinfo {author} {\bibfnamefont {O.}~\bibnamefont {Astafiev}},\ }\href
  {https://doi.org/10.1063/5.0011746} {\bibfield  {journal} {\bibinfo
  {journal} {AIP Conference Proceedings}\ }\textbf {\bibinfo {volume} {2241}},\
  \bibinfo {pages} {020036} (\bibinfo {year} {2020})}\BibitemShut {NoStop}%
\bibitem [{\citenamefont {Gunin}\ \emph {et~al.}(2023)\citenamefont {Gunin},
  \citenamefont {Dmitriev}, \citenamefont {Vasenin}, \citenamefont {Tikhonov},
  \citenamefont {Fedorov},\ and\ \citenamefont {Astafiev}}]{Gunin2023}%
  \BibitemOpen
  \bibfield  {author} {\bibinfo {author} {\bibfnamefont {S.~A.}\ \bibnamefont
  {Gunin}}, \bibinfo {author} {\bibfnamefont {A.~Y.}\ \bibnamefont {Dmitriev}},
  \bibinfo {author} {\bibfnamefont {A.~V.}\ \bibnamefont {Vasenin}}, \bibinfo
  {author} {\bibfnamefont {K.~S.}\ \bibnamefont {Tikhonov}}, \bibinfo {author}
  {\bibfnamefont {G.~P.}\ \bibnamefont {Fedorov}},\ and\ \bibinfo {author}
  {\bibfnamefont {O.~V.}\ \bibnamefont {Astafiev}},\ }\href
  {https://doi.org/10.1103/PhysRevA.108.033723} {\bibfield  {journal} {\bibinfo
   {journal} {Phys. Rev. A}\ }\textbf {\bibinfo {volume} {108}},\ \bibinfo
  {pages} {033723} (\bibinfo {year} {2023})}\BibitemShut {NoStop}%
\bibitem [{\citenamefont {Kolobov}\ and\ \citenamefont
  {Sokolov}(1987)}]{kolobov1987quantum}%
  \BibitemOpen
  \bibfield  {author} {\bibinfo {author} {\bibfnamefont {M.}~\bibnamefont
  {Kolobov}}\ and\ \bibinfo {author} {\bibfnamefont {I.}~\bibnamefont
  {Sokolov}},\ }\href@noop {} {\bibfield  {journal} {\bibinfo  {journal}
  {Optics and Spectroscopy}\ }\textbf {\bibinfo {volume} {62}},\ \bibinfo
  {pages} {69} (\bibinfo {year} {1987})}\BibitemShut {NoStop}%
\bibitem [{\citenamefont {Gardiner}(1993)}]{gardiner1993driving}%
  \BibitemOpen
  \bibfield  {author} {\bibinfo {author} {\bibfnamefont {C.}~\bibnamefont
  {Gardiner}},\ }\href {https://doi.org/10.1103/PhysRevLett.70.2269} {\bibfield
   {journal} {\bibinfo  {journal} {Phys.~Rev.~Lett.}\ }\textbf {\bibinfo
  {volume} {70}},\ \bibinfo {pages} {2269} (\bibinfo {year}
  {1993})}\BibitemShut {NoStop}%
\bibitem [{\citenamefont {Pfaff}\ \emph {et~al.}(2014)\citenamefont {Pfaff},
  \citenamefont {Hensen}, \citenamefont {Bernien}, \citenamefont {van Dam},
  \citenamefont {Blok}, \citenamefont {Taminiau}, \citenamefont {Tiggelman},
  \citenamefont {Schouten}, \citenamefont {Markham}, \citenamefont {Twitchen}
  \emph {et~al.}}]{pfaff2014unconditional}%
  \BibitemOpen
  \bibfield  {author} {\bibinfo {author} {\bibfnamefont {W.}~\bibnamefont
  {Pfaff}}, \bibinfo {author} {\bibfnamefont {B.~J.}\ \bibnamefont {Hensen}},
  \bibinfo {author} {\bibfnamefont {H.}~\bibnamefont {Bernien}}, \bibinfo
  {author} {\bibfnamefont {S.~B.}\ \bibnamefont {van Dam}}, \bibinfo {author}
  {\bibfnamefont {M.~S.}\ \bibnamefont {Blok}}, \bibinfo {author}
  {\bibfnamefont {T.~H.}\ \bibnamefont {Taminiau}}, \bibinfo {author}
  {\bibfnamefont {M.~J.}\ \bibnamefont {Tiggelman}}, \bibinfo {author}
  {\bibfnamefont {R.~N.}\ \bibnamefont {Schouten}}, \bibinfo {author}
  {\bibfnamefont {M.}~\bibnamefont {Markham}}, \bibinfo {author} {\bibfnamefont
  {D.~J.}\ \bibnamefont {Twitchen}}, \emph {et~al.},\ }\href
  {https://www.science.org/doi/full/10.1126/science.1253512} {\bibfield
  {journal} {\bibinfo  {journal} {Science}\ }\textbf {\bibinfo {volume}
  {345}},\ \bibinfo {pages} {532} (\bibinfo {year} {2014})}\BibitemShut
  {NoStop}%
\bibitem [{\citenamefont {Meyer}\ \emph {et~al.}(2015)\citenamefont {Meyer},
  \citenamefont {Stockill}, \citenamefont {Steiner}, \citenamefont {Le~Gall},
  \citenamefont {Matthiesen}, \citenamefont {Clarke}, \citenamefont {Ludwig},
  \citenamefont {Reichel}, \citenamefont {Atat\"ure},\ and\ \citenamefont
  {K\"ohl}}]{Meyer2015}%
  \BibitemOpen
  \bibfield  {author} {\bibinfo {author} {\bibfnamefont {H.~M.}\ \bibnamefont
  {Meyer}}, \bibinfo {author} {\bibfnamefont {R.}~\bibnamefont {Stockill}},
  \bibinfo {author} {\bibfnamefont {M.}~\bibnamefont {Steiner}}, \bibinfo
  {author} {\bibfnamefont {C.}~\bibnamefont {Le~Gall}}, \bibinfo {author}
  {\bibfnamefont {C.}~\bibnamefont {Matthiesen}}, \bibinfo {author}
  {\bibfnamefont {E.}~\bibnamefont {Clarke}}, \bibinfo {author} {\bibfnamefont
  {A.}~\bibnamefont {Ludwig}}, \bibinfo {author} {\bibfnamefont
  {J.}~\bibnamefont {Reichel}}, \bibinfo {author} {\bibfnamefont
  {M.}~\bibnamefont {Atat\"ure}},\ and\ \bibinfo {author} {\bibfnamefont
  {M.}~\bibnamefont {K\"ohl}},\ }\href
  {https://doi.org/10.1103/PhysRevLett.114.123001} {\bibfield  {journal}
  {\bibinfo  {journal} {Phys. Rev. Lett.}\ }\textbf {\bibinfo {volume} {114}},\
  \bibinfo {pages} {123001} (\bibinfo {year} {2015})}\BibitemShut {NoStop}%
\bibitem [{\citenamefont {Delteil}\ \emph {et~al.}(2017)\citenamefont
  {Delteil}, \citenamefont {Sun}, \citenamefont {F\"alt},\ and\ \citenamefont
  {Imamo\ifmmode~\breve{g}\else \u{g}\fi{}lu}}]{Delteil2017}%
  \BibitemOpen
  \bibfield  {author} {\bibinfo {author} {\bibfnamefont {A.}~\bibnamefont
  {Delteil}}, \bibinfo {author} {\bibfnamefont {Z.}~\bibnamefont {Sun}},
  \bibinfo {author} {\bibfnamefont {S.}~\bibnamefont {F\"alt}},\ and\ \bibinfo
  {author} {\bibfnamefont {A.}~\bibnamefont {Imamo\ifmmode~\breve{g}\else
  \u{g}\fi{}lu}},\ }\href {https://doi.org/10.1103/PhysRevLett.118.177401}
  {\bibfield  {journal} {\bibinfo  {journal} {Phys. Rev. Lett.}\ }\textbf
  {\bibinfo {volume} {118}},\ \bibinfo {pages} {177401} (\bibinfo {year}
  {2017})}\BibitemShut {NoStop}%
\bibitem [{\citenamefont {Peng}\ \emph {et~al.}(2016)\citenamefont {Peng},
  \citenamefont {De~Graaf}, \citenamefont {Tsai},\ and\ \citenamefont
  {Astafiev}}]{peng2016tuneable}%
  \BibitemOpen
  \bibfield  {author} {\bibinfo {author} {\bibfnamefont {Z.}~\bibnamefont
  {Peng}}, \bibinfo {author} {\bibfnamefont {S.}~\bibnamefont {De~Graaf}},
  \bibinfo {author} {\bibfnamefont {J.}~\bibnamefont {Tsai}},\ and\ \bibinfo
  {author} {\bibfnamefont {O.}~\bibnamefont {Astafiev}},\ }\href
  {https://www.nature.com/articles/ncomms12588} {\bibfield  {journal} {\bibinfo
   {journal} {Nature Communications}\ }\textbf {\bibinfo {volume} {7}},\
  \bibinfo {pages} {12588} (\bibinfo {year} {2016})}\BibitemShut {NoStop}%
\bibitem [{\citenamefont {Hanschke}\ \emph {et~al.}(2020)\citenamefont
  {Hanschke}, \citenamefont {Schweickert}, \citenamefont {Carre\~no},
  \citenamefont {Sch\"oll}, \citenamefont {Zeuner}, \citenamefont {Lettner},
  \citenamefont {Casalengua}, \citenamefont {Reindl}, \citenamefont {da~Silva},
  \citenamefont {Trotta}, \citenamefont {Finley}, \citenamefont {Rastelli},
  \citenamefont {del Valle}, \citenamefont {Laussy}, \citenamefont {Zwiller},
  \citenamefont {M\"uller},\ and\ \citenamefont
  {J\"ons}}]{PhysRevLett.125.170402}%
  \BibitemOpen
  \bibfield  {author} {\bibinfo {author} {\bibfnamefont {L.}~\bibnamefont
  {Hanschke}}, \bibinfo {author} {\bibfnamefont {L.}~\bibnamefont
  {Schweickert}}, \bibinfo {author} {\bibfnamefont {J.~C.~L.}\ \bibnamefont
  {Carre\~no}}, \bibinfo {author} {\bibfnamefont {E.}~\bibnamefont {Sch\"oll}},
  \bibinfo {author} {\bibfnamefont {K.~D.}\ \bibnamefont {Zeuner}}, \bibinfo
  {author} {\bibfnamefont {T.}~\bibnamefont {Lettner}}, \bibinfo {author}
  {\bibfnamefont {E.~Z.}\ \bibnamefont {Casalengua}}, \bibinfo {author}
  {\bibfnamefont {M.}~\bibnamefont {Reindl}}, \bibinfo {author} {\bibfnamefont
  {S.~F.~C.}\ \bibnamefont {da~Silva}}, \bibinfo {author} {\bibfnamefont
  {R.}~\bibnamefont {Trotta}}, \bibinfo {author} {\bibfnamefont {J.~J.}\
  \bibnamefont {Finley}}, \bibinfo {author} {\bibfnamefont {A.}~\bibnamefont
  {Rastelli}}, \bibinfo {author} {\bibfnamefont {E.}~\bibnamefont {del Valle}},
  \bibinfo {author} {\bibfnamefont {F.~P.}\ \bibnamefont {Laussy}}, \bibinfo
  {author} {\bibfnamefont {V.}~\bibnamefont {Zwiller}}, \bibinfo {author}
  {\bibfnamefont {K.}~\bibnamefont {M\"uller}},\ and\ \bibinfo {author}
  {\bibfnamefont {K.~D.}\ \bibnamefont {J\"ons}},\ }\href
  {https://doi.org/10.1103/PhysRevLett.125.170402} {\bibfield  {journal}
  {\bibinfo  {journal} {Phys. Rev. Lett.}\ }\textbf {\bibinfo {volume} {125}},\
  \bibinfo {pages} {170402} (\bibinfo {year} {2020})}\BibitemShut {NoStop}%
\bibitem [{\citenamefont {Kimble}\ \emph
  {et~al.}(1977{\natexlab{b}})\citenamefont {Kimble}, \citenamefont
  {Dagenais},\ and\ \citenamefont {Mandel}}]{Kimble1977}%
  \BibitemOpen
  \bibfield  {author} {\bibinfo {author} {\bibfnamefont {H.~J.}\ \bibnamefont
  {Kimble}}, \bibinfo {author} {\bibfnamefont {M.}~\bibnamefont {Dagenais}},\
  and\ \bibinfo {author} {\bibfnamefont {L.}~\bibnamefont {Mandel}},\ }\href
  {https://doi.org/10.1103/PhysRevLett.39.691} {\bibfield  {journal} {\bibinfo
  {journal} {Phys. Rev. Lett.}\ }\textbf {\bibinfo {volume} {39}},\ \bibinfo
  {pages} {691} (\bibinfo {year} {1977}{\natexlab{b}})}\BibitemShut {NoStop}%
\bibitem [{\citenamefont {Mandel}(1979)}]{Mandel:79}%
  \BibitemOpen
  \bibfield  {author} {\bibinfo {author} {\bibfnamefont {L.}~\bibnamefont
  {Mandel}},\ }\href {https://doi.org/10.1364/OL.4.000205} {\bibfield
  {journal} {\bibinfo  {journal} {Opt. Lett.}\ }\textbf {\bibinfo {volume}
  {4}},\ \bibinfo {pages} {205} (\bibinfo {year} {1979})}\BibitemShut {NoStop}%
\bibitem [{\citenamefont {Treussart}\ \emph {et~al.}(2002)\citenamefont
  {Treussart}, \citenamefont {All\'eaume}, \citenamefont {Le~Floc'h},
  \citenamefont {Xiao}, \citenamefont {Courty},\ and\ \citenamefont
  {Roch}}]{Treussart2002direct}%
  \BibitemOpen
  \bibfield  {author} {\bibinfo {author} {\bibfnamefont {F.}~\bibnamefont
  {Treussart}}, \bibinfo {author} {\bibfnamefont {R.}~\bibnamefont
  {All\'eaume}}, \bibinfo {author} {\bibfnamefont {V.}~\bibnamefont
  {Le~Floc'h}}, \bibinfo {author} {\bibfnamefont {L.~T.}\ \bibnamefont {Xiao}},
  \bibinfo {author} {\bibfnamefont {J.-M.}\ \bibnamefont {Courty}},\ and\
  \bibinfo {author} {\bibfnamefont {J.-F.}\ \bibnamefont {Roch}},\ }\href
  {https://doi.org/10.1103/PhysRevLett.89.093601} {\bibfield  {journal}
  {\bibinfo  {journal} {Phys. Rev. Lett.}\ }\textbf {\bibinfo {volume} {89}},\
  \bibinfo {pages} {093601} (\bibinfo {year} {2002})}\BibitemShut {NoStop}%
\bibitem [{\citenamefont {L\'opez Carre\~no}\ \emph {et~al.}(2022)\citenamefont
  {L\'opez Carre\~no}, \citenamefont {Zubizarreta~Casalengua}, \citenamefont
  {Silva}, \citenamefont {del Valle},\ and\ \citenamefont
  {Laussy}}]{Carreno2022Loss}%
  \BibitemOpen
  \bibfield  {author} {\bibinfo {author} {\bibfnamefont {J.~C.}\ \bibnamefont
  {L\'opez Carre\~no}}, \bibinfo {author} {\bibfnamefont {E.}~\bibnamefont
  {Zubizarreta~Casalengua}}, \bibinfo {author} {\bibfnamefont {B.}~\bibnamefont
  {Silva}}, \bibinfo {author} {\bibfnamefont {E.}~\bibnamefont {del Valle}},\
  and\ \bibinfo {author} {\bibfnamefont {F.~P.}\ \bibnamefont {Laussy}},\
  }\href {https://doi.org/10.1103/PhysRevA.105.023724} {\bibfield  {journal}
  {\bibinfo  {journal} {Phys. Rev. A}\ }\textbf {\bibinfo {volume} {105}},\
  \bibinfo {pages} {023724} (\bibinfo {year} {2022})}\BibitemShut {NoStop}%
\end{thebibliography}%

\end{document}